%% file: paper.tex
\documentclass[12pt,a4paper,dvips]{article}
\usepackage{a4p}
\usepackage{cite,mcite}
\usepackage{graphicx}
\usepackage{physics }
\usepackage{l3_title,ifthen}
\usepackage{amssymb}
\journalname{Phys. Lett. B}
\date{October 25, 2002}

\preprint{2002-080}
\newcommand{\EE}{\ensuremath{\mathrm{e}^+\mathrm{e}^-}}
\newcommand{\WW}{\ensuremath{\mathrm{W}^+\mathrm{W}^-}}
\newcommand{\ZZ}{\ensuremath{\mathrm{ZZ}}}
\newcommand{\QQ}{\ensuremath{\mathrm{q}\bar{\mathrm{q}}}}

\newcommand{\htogg}{\ensuremath{\mathrm{h}\rightarrow\gamma\gamma}}
\newcommand{\htoWW}{\ensuremath{\mathrm{h}\rightarrow\mathrm{WW}^*}}
\newcommand{\htoZZ}{\ensuremath{\mathrm{h}\rightarrow\mathrm{ZZ}^*}}

\newcommand{\HWW}{\ensuremath{\mathrm{h}\rightarrow\mathrm{WW}^*}}
\newcommand{\HZZ}{\ensuremath{\mathrm{h}\rightarrow\mathrm{ZZ}^*}}

\newcommand{\mh}{\ensuremath{m_\mathrm{h}}}
\newcommand{\mW}{\ensuremath{m_\mathrm{W}}}
\newcommand{\mZ}{\ensuremath{m_\mathrm{Z}}}

\newcommand{\Brgg}{\ensuremath{R_{\gamma\gamma}}}
\newcommand{\Brphobic}{\ensuremath{\mathrm{BR}_\mathrm{bosons}}}
\newcommand{\qqqqlv}{\ensuremath{\mathrm{qqqq}\ell\nu}}

\newlength{\capindent}
\newlength{\capwidth}
\newlength{\figwidth}
\newcommand{\icaption}[2][!*!,!]{\hspace*{\capindent}
  \begin{minipage}{\capwidth}
    \ifthenelse{\equal{#1}{!*!,!}}
      {\caption{#2}}
      {\caption[#1]{#2}}
  \end{minipage}}
\newcommand{\pho}{\phantom{0}}

\begin{document}
\begin{titlepage}
\title{ 
Search for a Higgs Boson\\Decaying to Weak Boson Pairs\\at LEP}
\author{The L3 Collaboration}
%
% The abstract
%

\begin{abstract} 

A Higgs particle produced in association with a Z boson and decaying
into weak boson pairs is searched for in 336.4 pb$^{-1}$ of data
collected by the L3 experiment at LEP at centre-of-mass energies from
200 to 209 \GeV.  Limits on the branching fraction of the Higgs boson
decay into two weak bosons as a function of the Higgs mass are
derived.  These results are combined with the L3 search for a Higgs
boson decaying to photon pairs.  A Higgs produced with a Standard
Model $\EE\to\mathrm{Zh}$ cross section and decaying only into
electroweak boson pairs is excluded at 95\% CL for a mass below $ 107
\GeV$.

\end{abstract} 
\submitted 
\end{titlepage} 

\section{Introduction}

In the Standard Model, the Higgs mechanism generates the masses of
elementary particles and stabilises the high energy behaviour of the
electroweak interactions. However, the Standard Model does not predict
the mass of the Higgs boson. Searches to date have focused on b quark
decays of the Higgs, but the model predicts an increased branching
fraction to massive vector boson pairs for a heavier Higgs. In some
extensions of the Standard Model which predict multiple Higgs bosons,
the lightest Higgs boson couples primarily to bosons, not fermions
\cite{om}. Results excluding these ``fermiophobic'' models were
reported for Higgs decays into two photons, but for Higgs masses above
90$\GeV$, the decay into massive vector boson pairs
dominates\cite{fphob}.

\begin{figure}[H]
\begin{center}
\includegraphics[width=12cm]{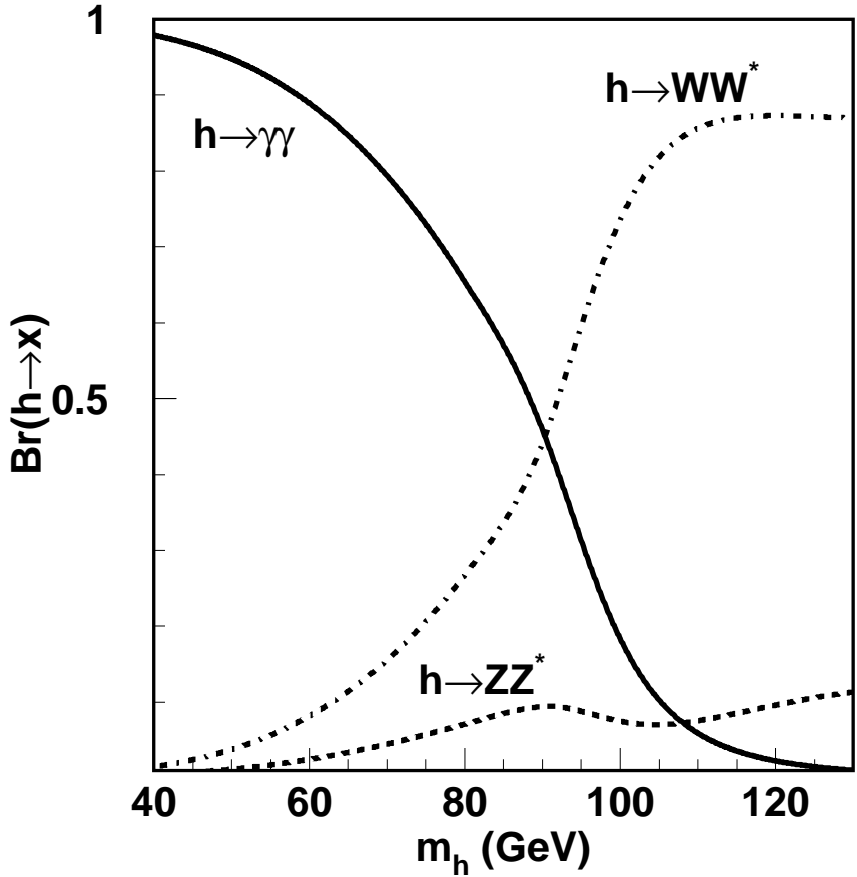} 
\caption{
Higgs branching fractions predicted in the benchmark fermiophobic scenario.
}
\label{fig_phobebr}
\end{center}
\end{figure}

Several possible models predict the presence of a fermiophobic
Higgs. As a point of reference, a simple extension of the Standard
Model is chosen where the production cross section for
$\EE\to\mathrm{Zh}$, the Higgsstrahlung process, is kept at the
Standard Model value.  All direct decays into fermions are removed,
and the resulting branching ratios favour decays into two photons for
Higgs bosons with masses below 90 \GeV, while massive vector boson
pairs become important above 90 \GeV. This set of assumptions is
called the ``benchmark fermiophobic scenario''.  The branching ratios
to $\gamma\gamma$, $\WW$, and $\ZZ$ predicted in this scenario are
plotted in Figure~\ref{fig_phobebr}.

This Letter presents the search for a Higgs boson produced in
association with a Z boson through the process $\EE\to\mathrm{Zh}$,
followed by the decay of the Higgs to either a pair of W or Z bosons.
Throughout the analysis we assume that the ratio between the branching
fraction \HWW\ and the branching fraction \HZZ\ is given by the
Standard Model expectations. This assumption is valid for scenarios
beyond the Standard Model satisfying the constraint $\rho\approx 1$
\cite{cite_wwzz}. Consequently, the branching fraction \HWW\ is
expected to be the dominant weak boson decay for all Higgs masses.
Accordingly, the analyses focus on \HWW\, but the efficiency of the
analyses for a \HZZ\ signal is considered where appropriate.  Since
the Higgsstrahlung mass reach of the LEP data is less than 160 \GeV,
at least one of the weak bosons produced from the Higgs decay must be
off its mass shell. The analyses assume the presence of one vector
boson with a mass within $\pm 10\GeV$ of its nominal mass and one with
a much smaller mass.  This topology is expected to be present in
approximately 70\% of signal events.

\section{Data and Monte Carlo Samples}

We analyse data collected with the L3 detector \cite{l3det} at
centre-of-mass energies $\sqrt{\mathrm{s}}=200-209\GeV$, for a total
integrated luminosity of 336.4 pb$^{-1}$. The data are grouped into
six samples whose average centre-of-mass energies and corresponding
integrated luminosities are listed in Table \ref{tab_lumi}.

The Higgs signal cross section and fermiophobic branching ratios are
calculated using the HZHA Monte Carlo generator \cite{hzha}. For
efficiency studies, Higgs events are generated using PYTHIA
\cite{pythia} for Higgs masses between 80$\GeV$ and 120$\GeV$, with a
mass step of 5$\GeV$ up to 100$\GeV$ and 1$\GeV$ thereafter.  The
$\EE\to\mathrm{ZZ}$, $\EE\to\mathrm{Z}\EE$, and $\EE\to\QQ(\gamma)$
processes are simulated with the PYTHIA generator.  The
KK2f\cite{kk2f} generator is also used for the $\EE\to\QQ(\gamma)$
process.  The KORALW \cite{koralw} generator is used for the
$\EE\to\WW$ process except for the
$\EE\to\mathrm{e}^-\bar{\nu}_\mathrm{e}\mathrm{q}\bar{\mathrm{q}}^\prime$
final state, which is simulated using EXCALIBUR\cite{excalibur}.
Hadron production in two-photon interactions is modelled using the
PHOJET program\cite{phojet}.

The L3 detector response is simulated using the GEANT
program\cite{geant}, which takes into account the effects of energy
loss, multiple scattering and showering in the detector.  The GHEISHA
package\cite{gheisha} is used to simulate hadronic interactions in the
detector.  Time dependent detector inefficiencies, as monitored during
the data acquisition period, are also simulated.

\section{Analyses}

Of the possible final states from the $\EE\to \mathrm{Zh}\to
\mathrm{ZWW^*} \to 6$ {\it fermions} process, this Letter describes
the results of analyses for six of them, covering 93\% of the total
branching ratio.  The six decay channels searched are:
$\mathrm{Zh}\to\mathrm{qqqqqq}$, $\mathrm{Zh}\to\mathrm{qqqq}\ell\nu$,
$\mathrm{Zh}\to\mathrm{qq}\ell\nu\ell\nu$,
$\mathrm{Zh}\to\nu\nu\mathrm{qqqq}$,
$\mathrm{Zh}\to\nu\nu\mathrm{qq}\ell\nu$, and
$\mathrm{Zh}\to\ell\ell\mathrm{qqqq}$~\footnote{This simplified
notation, covering all possible combinations of flavour and charge, is
adopted throughout this Letter.}.  The analyses adopt a common
procedure of preselection, followed by a selection using a neural
network approach and the production of a final discriminant variable.

The first step is to apply a set of preselection cuts to remove the
background events most different from the signal.  Important
preselection cuts are made on visible energy and counts of tracks and
clusters.  The number of neutrinos in a channel determines the visible
energy window, while the number of jets provides a basis for cuts on
the number of charged tracks and calorimeter clusters.  For channels
containing leptons, important lepton identification cuts are applied.
Two-photon events and $\EE\to\QQ(\gamma)$ events containing one or
more photons from initial state radiation are suppressed using cuts on
the fraction of $\sqrt{s}$ deposited in detectors near the beam line
and by cutting events where the missing momentum vector points near
the beam pipe.  These cuts also reduce the number of
$\EE\to\mathrm{e}\nu\mathrm{qq}$ background events accepted by the
analysis.

The background processes which are more similar to the signal are
removed using one or more neural networks~\cite{snns}.  Where
possible, preselected events are subjected to a constrained fit to the
expected event topology before constructing the neural
networks. Variables considered in the networks include the energies of
the most and least energetic jets, the minimum angle between any two
jets, the minimum angle between any jet and any lepton, and the
reconstructed masses of any Z or W bosons identified in the event.

After applying cuts on the neural networks, the analyses produce final
distributions of the selected signal, background, and data, generally
using a discriminant combination of the neural network outputs and a
reconstructed Higgs mass.

In the following sections we outline the selection procedures for the
six channels.  Complete details are available in
Reference~\citen{thesis}.

\subsection{\boldmath{$\mathrm{Zh}\to\mathrm{qqqqqq}$}}

The qqqqqq analysis searches for the case when the W and Z bosons
decay into hadrons.  The primary backgrounds to this channel are
$\mathrm{ZZ}\to\mathrm{qqqq}$, $\WW\to\mathrm{qqqq}$ and
$\EE\to\QQ(\gamma)$ events. This six-jet signature is also produced by
the $\mathrm{Zh}\rightarrow\mathrm{ZZZ}^{*}\rightarrow\mathrm{qqqqqq}$
process, which represents 34\% of the total \HZZ\ signal.  The
efficiency for the \HZZ\ signal is included in the analysis,
effectively adding 15$\%$ to the expected rate relative to using only
\HWW .

Events with full energy and large hadronic content are selected.  The
analysis uses a fit of the event to a six-jet topology using the
Durham algorithm \cite{durham}. Events with a poor match to the
six-jet hypothesis are rejected.

After preselection, each six-jet event is subjected to a kinematic
fit, which requires momentum and energy conservation. Following the
fit, the pair of jets with invariant mass closest to \mZ\ is chosen as
the Z candidate. Of the remaining four jets, the dijet pair with
invariant mass closest to \mW\ is assigned to be the W candidate and
the remaining pair is identified as the $\mathrm{W}^*$.

 The analysis uses three neural networks, one for each of the
$\EE\to\WW$, $\EE\to\ZZ$, and the $\EE\to\QQ(\gamma)$ backgrounds. All
three neural networks use the same eleven inputs and have twenty
hidden nodes and one output node. The input variables are chosen
according to the specific features which differentiate signal from
background:

\begin{itemize}
\item Typical QCD features in the background are: unequal repartition
of the energy among the six jets, jets with low multiplicity, small
invariant masses and small angles between closer jets. The following
six input variables are therefore chosen: the energies of the most and
least energetic jets, $E^{max}_{jet}$ and $E^{min}_{jet}$, the minimum
number of tracks in a jet, $n^{min}_{jet}$, the minimum angle between
jets, $\theta^{min}_{jj}$ and the values of the Durham algorithm
parameter $\log y_{45}$ and $\log y_{56}$. The value $y_{mn}$ is the
threshold for which the DURHAM algorithm reconstructs a $m$-jet event
as a $n$-jet one.
\item The event must be kinematically inconsistent with the production
of only two on-shell boson pairs. This leads to the choice of the
following five input variables: the chi-square of a kinematic fit to
equal-mass boson pair-production, $\chi^2_{\W\W}$, the mass obtained in
the fit, $m^{fit}_{eq}$, the masses of the $\Zo$ and $\W$ candidates
after kinematic fit to the signal hypothesis, $m^{fit}_\Zo$ and
$m^{fit}_\W$, and the angle between the decay planes of the $\W$ and
$\W^{*}$ candidates, $\alpha_{\W\W^*}$.
\end{itemize}

\begin{figure}[H]
\begin{center}
\begin{tabular}{cc}
\multicolumn{2}{c}{\hspace{0.8cm}\includegraphics[width=13.5cm]{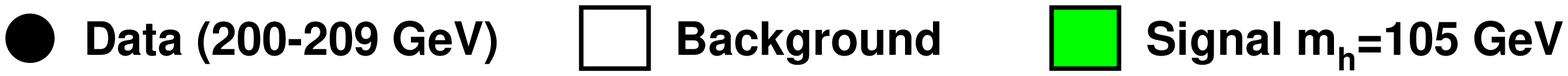}} \\
\includegraphics[width=7.00cm]{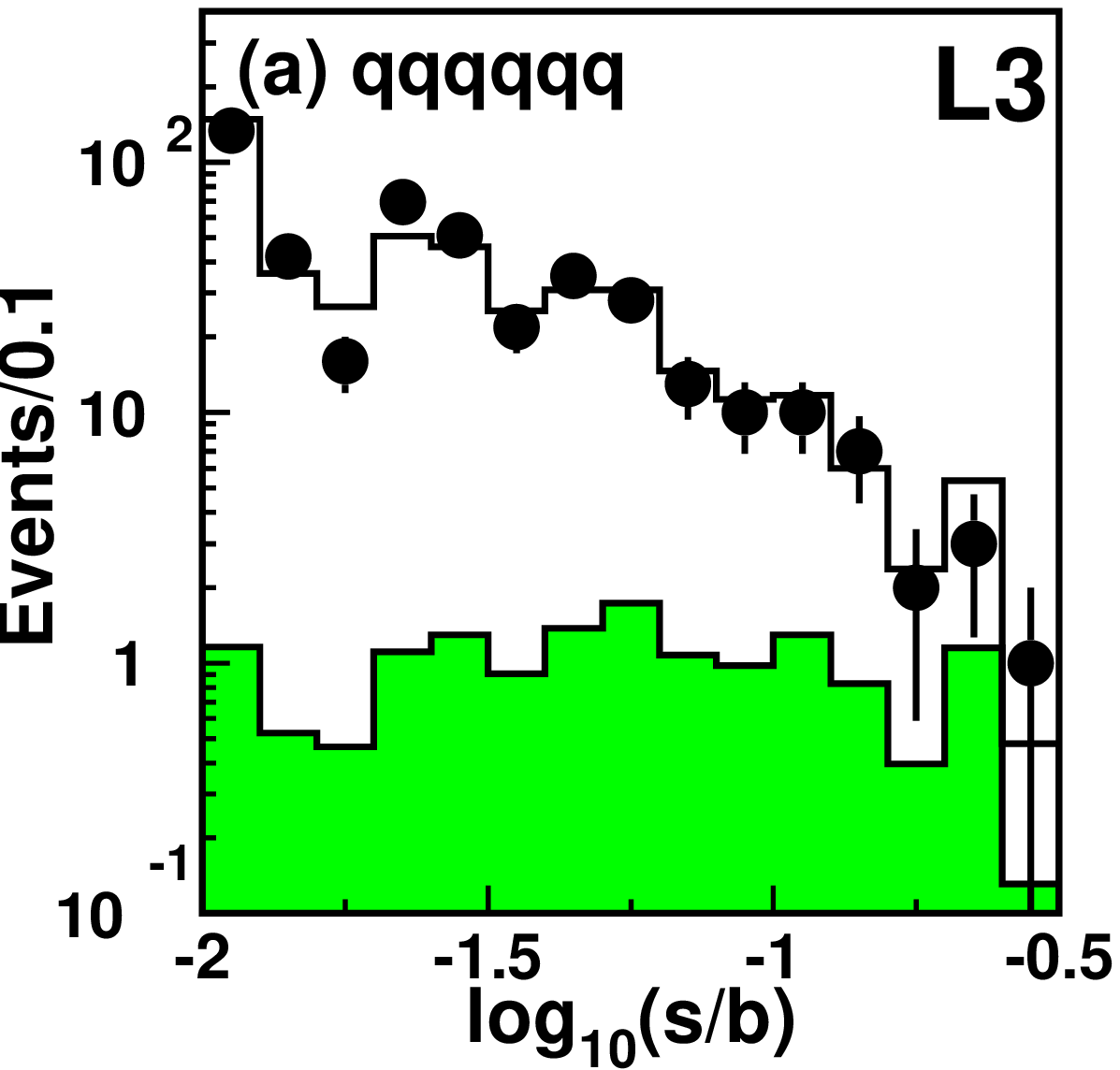} & \includegraphics[width=7.00cm]{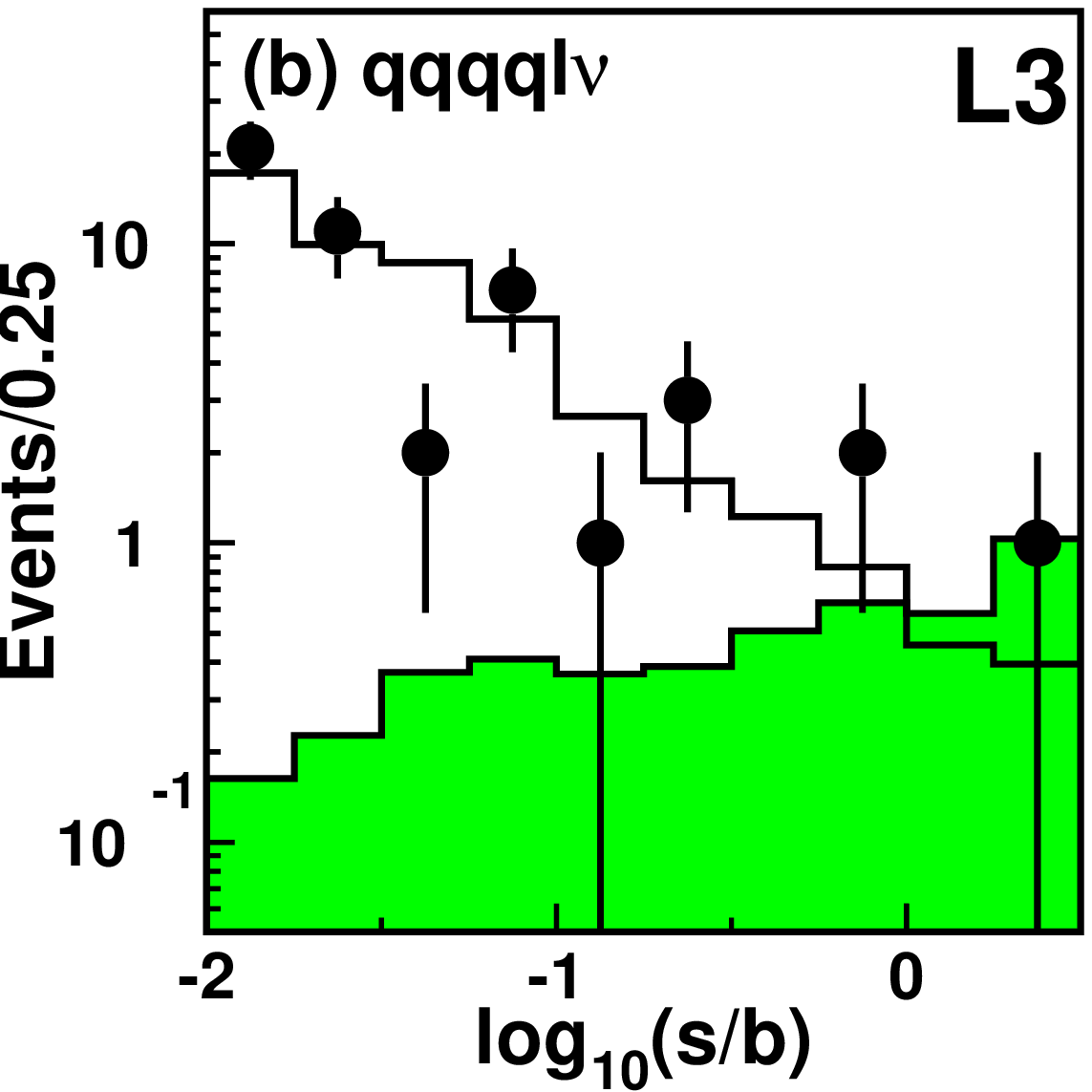} \\
\includegraphics[width=7.00cm]{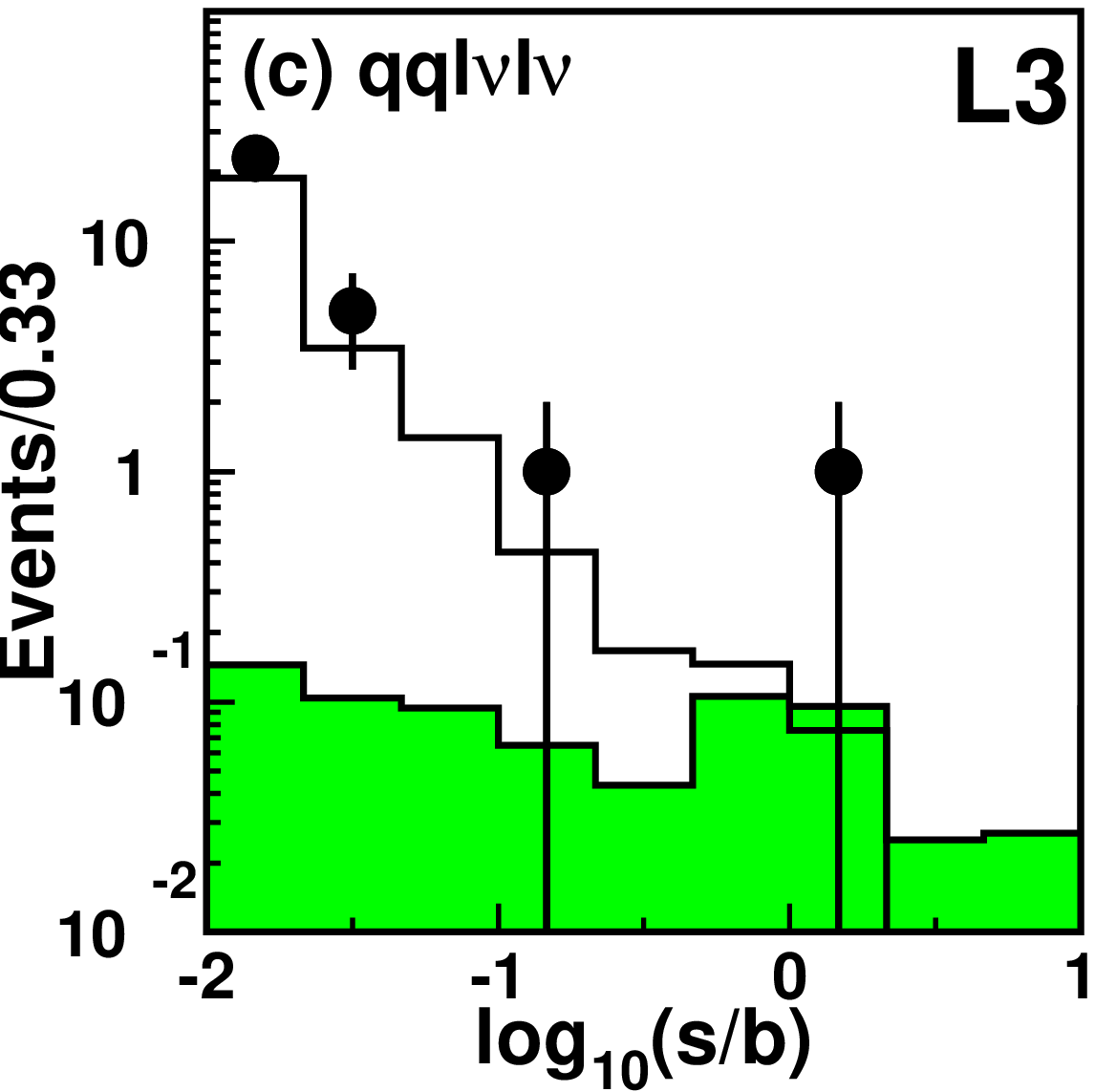} & \includegraphics[width=7.00cm]{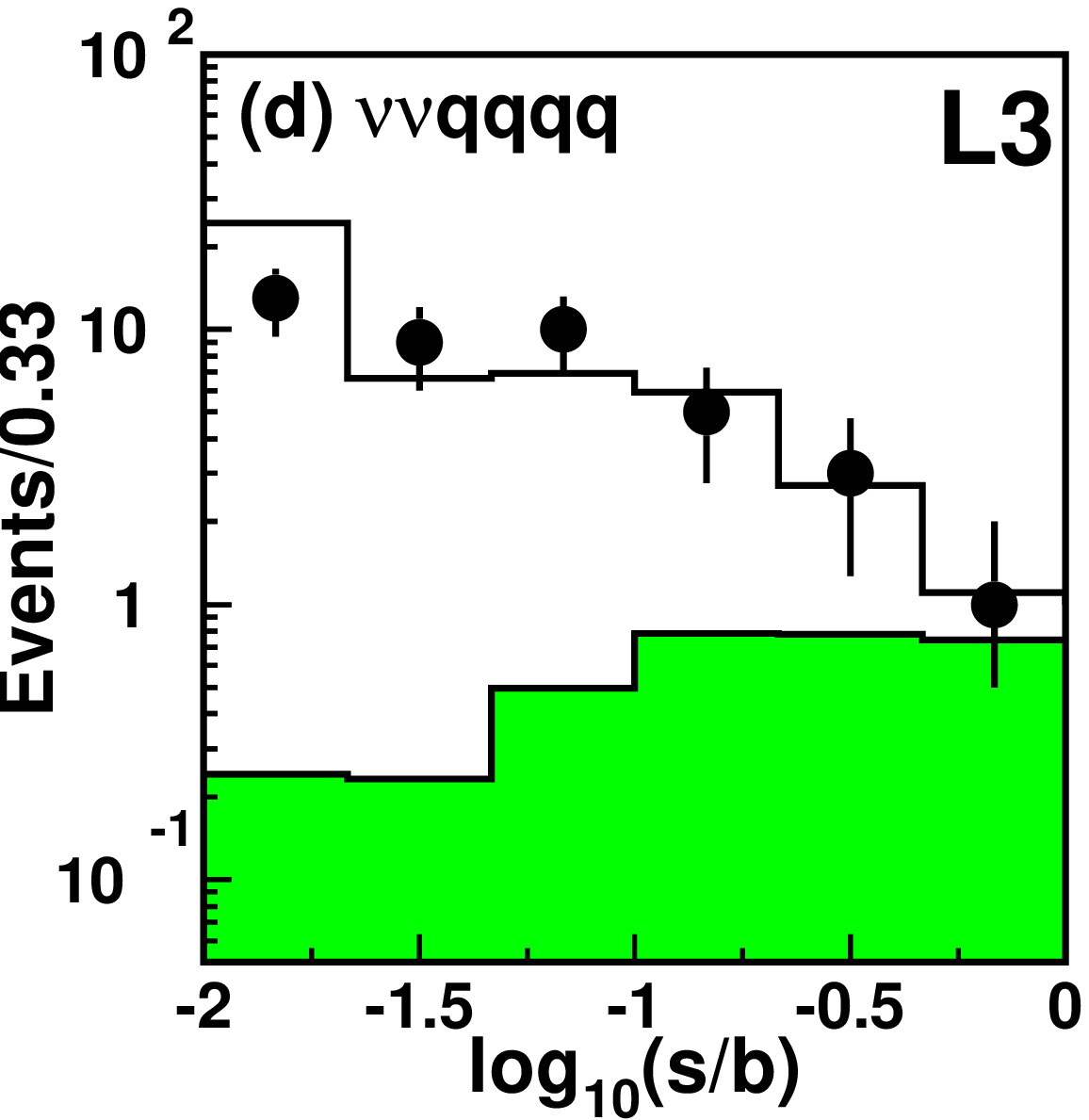} \\
\includegraphics[width=7.00cm]{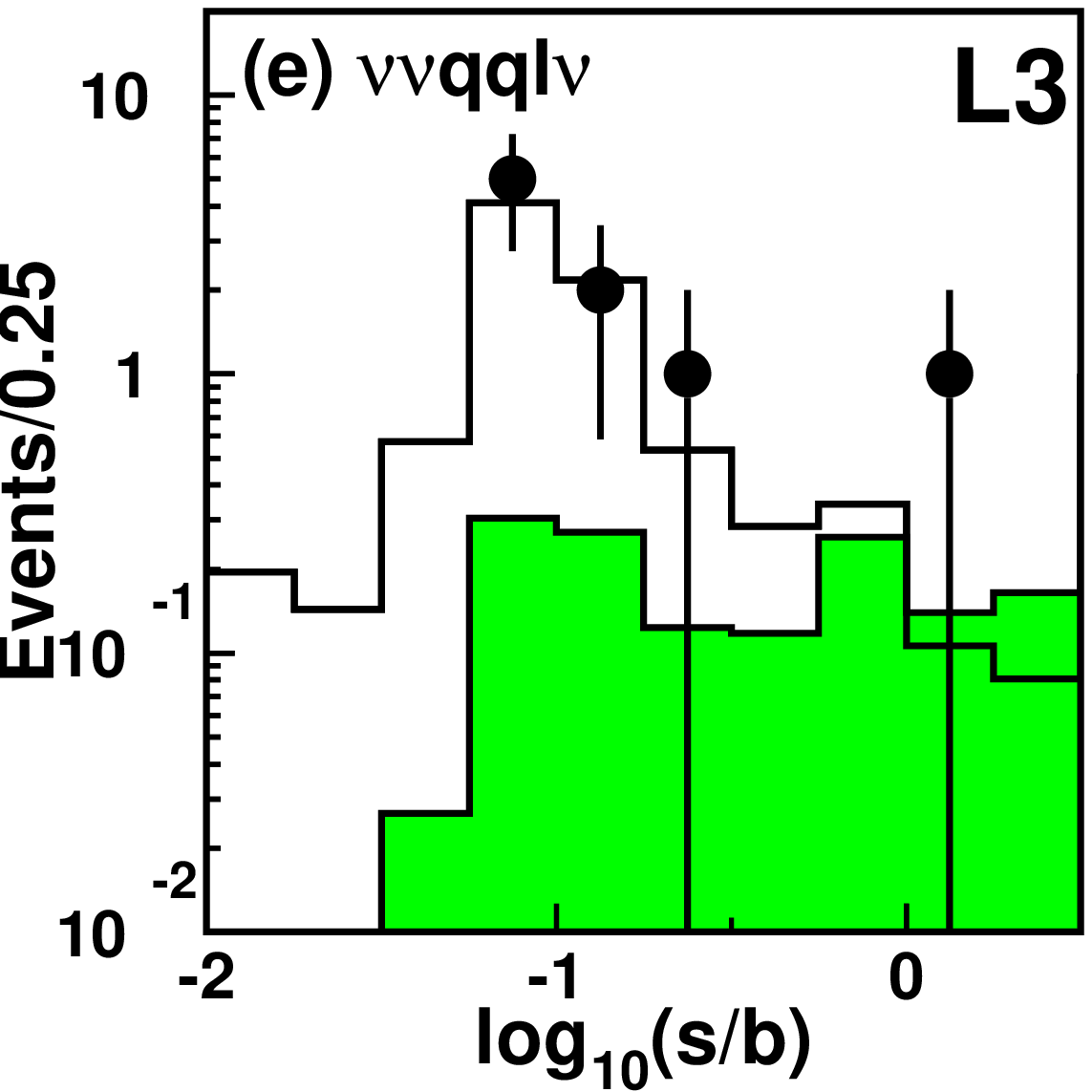} & \includegraphics[width=7.00cm]{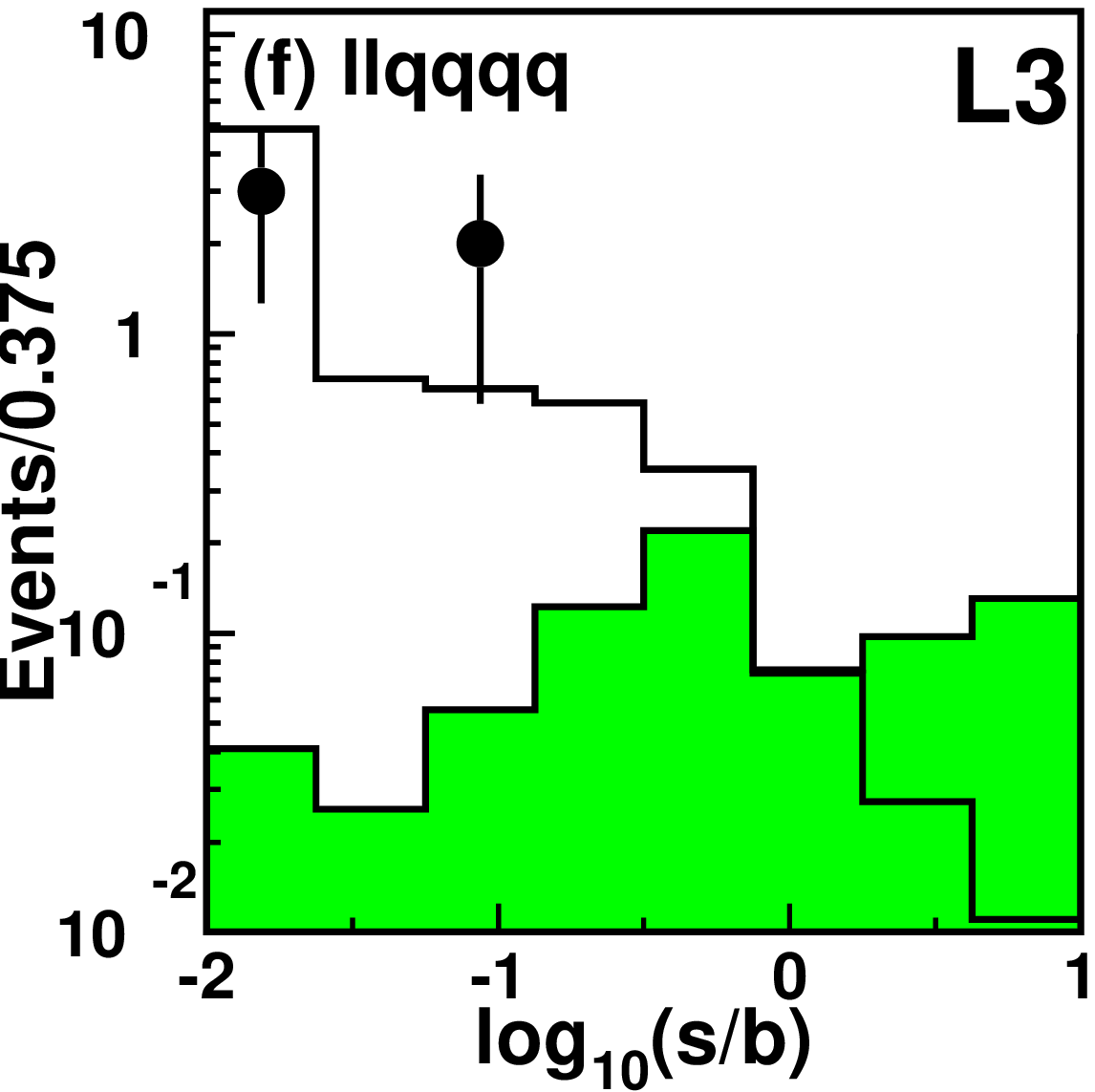} 
\end{tabular}
\caption{
Distributions of the final variable for data, background Monte Carlo and 
signal Monte Carlo at $\mh=105\GeV$.  All centre-of-mass energies and 
subchannels are combined.  The final variable is displayed in bins of 
$\log_{10}(\frac{\mathrm{signal}}{\mathrm{background}})$.
}
\label{fig_fv}
\end{center}
\end{figure}

As an example of the input variables to the neural networks,
Figure~\ref{fig_input}a presents the distribution of
$\theta^{min}_{jj}$ for  data, background and signal Monte Carlo.
Table \ref{tab_6j} gives the numbers of signal and background events
expected and data observed after cuts on the neural network output.
The final variable produced by the analysis is a discriminant
combining the three network outputs and the reconstructed Higgs mass
after the kinematic fit and Z candidate assignment.  Distributions of
this final variable for data, background and signal are plotted in
Figure~\ref{fig_fv}a.

\subsection{\boldmath{$\mathrm{Zh}\to\qqqqlv$}}

In the \qqqqlv\ channel, the Z decays into hadrons, while one W decays
into hadrons and the other decays into leptons. The different lepton
flavours naturally define three different subchannels:
$\mathrm{qqqqe}\nu$, $\mathrm{qqqq}\mu\nu$, and
$\mathrm{qqqq}\tau\nu$. Further, the difference between leptons coming
from the W and from the $\mathrm{W}^*$ doubles the number of
subchannels. In one set of signatures, the W decays into hadrons and
the $\mathrm{W}^*$ decays into leptons, which means the lepton and
neutrino energies are small, as is the missing energy of the event.
In the other set, the W decays into leptons and the $\mathrm{W}^*$
decays into hadrons, yielding a high energy lepton and large missing
energy. The major backgrounds are W pair production (especially
$\EE\to\WW\to\mathrm{qq}\tau\nu$), Z pair production, and the
$\EE\to\QQ(\gamma)$ process.

Events are classified into a subchannel according to the most
energetic lepton in the event. For the $\mathrm{qqqqe}\nu$ and
$\mathrm{qqqq}\mu\nu$ cases, the subchannels are separated using the
ratio of the lepton energy to the visible energy.  In the subchannels
where the lepton-neutrino pair comes from the on-shell W, both the
lepton and neutrino energies should be large.  For the subchannels
where the lepton and neutrino come from the W*, the converse is true.
In the $\mathrm{qqqq}\tau \nu$ channels, the initial lepton energy is
difficult to reconstruct, so the subchannels are separated using only
the visible energy.  The candidate lepton in the event is required to
meet minimum identification criteria to ensure isolated and well
reconstructed leptons.

 The analysis selects events with significant hadronic content and an
appropriate amount of visible energy depending on the subchannel. A
major source of background for this channel is the
$\WW\rightarrow\mathrm{qq}\ell\nu$ process.  To reduce it, an event is
rejected if its hadronic mass is less than 90 \GeV.  After
preselection, the major remaining background is $\EE\to\WW$,
particularly $\EE\to\WW\to \mathrm{qqqq}$ events with a lepton from
hadron decays.

\begin{figure}[H]
\begin{center}
\begin{tabular}{cc}
\includegraphics[width=7.00cm]{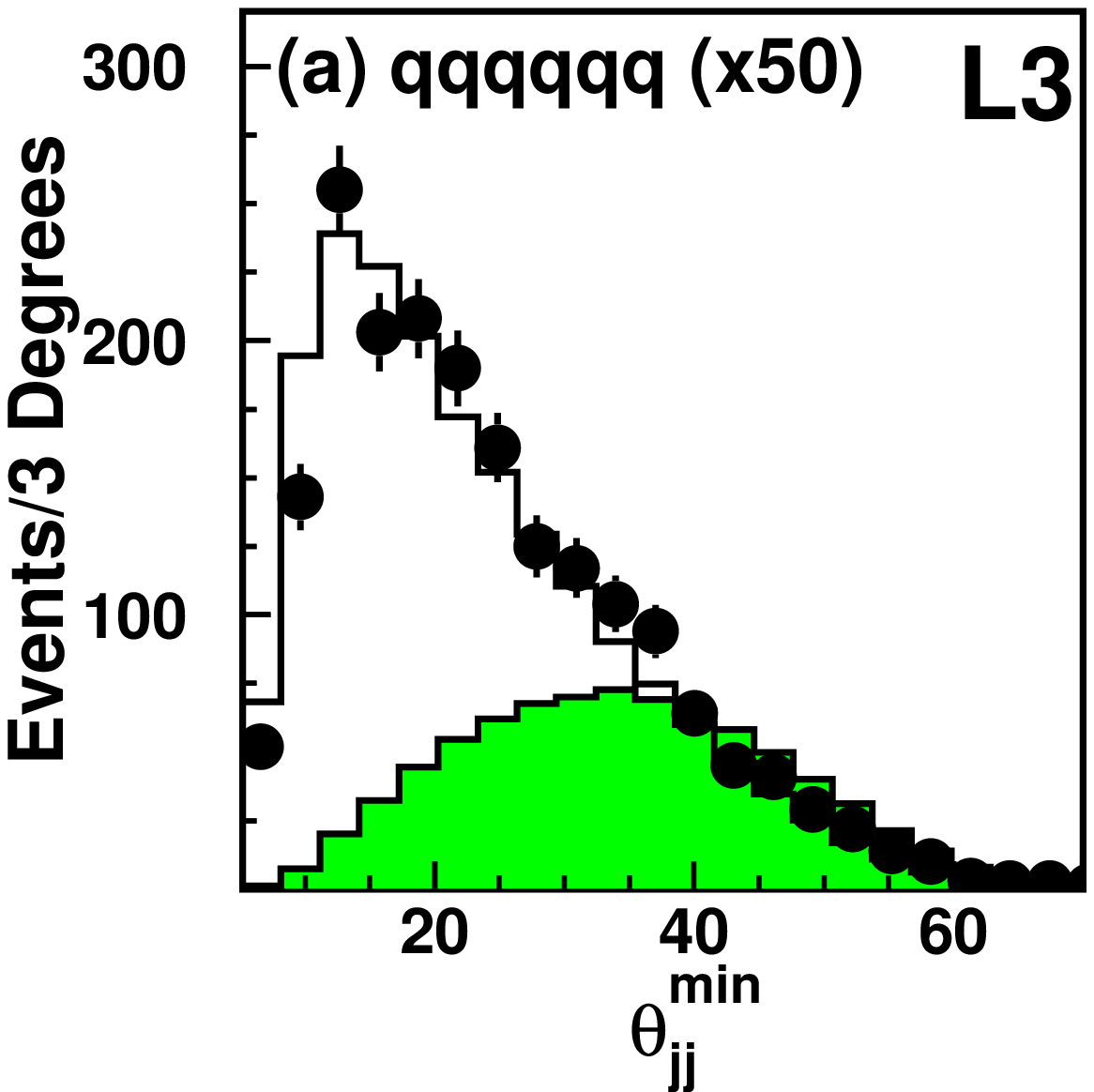} & \includegraphics[width=7.00cm]{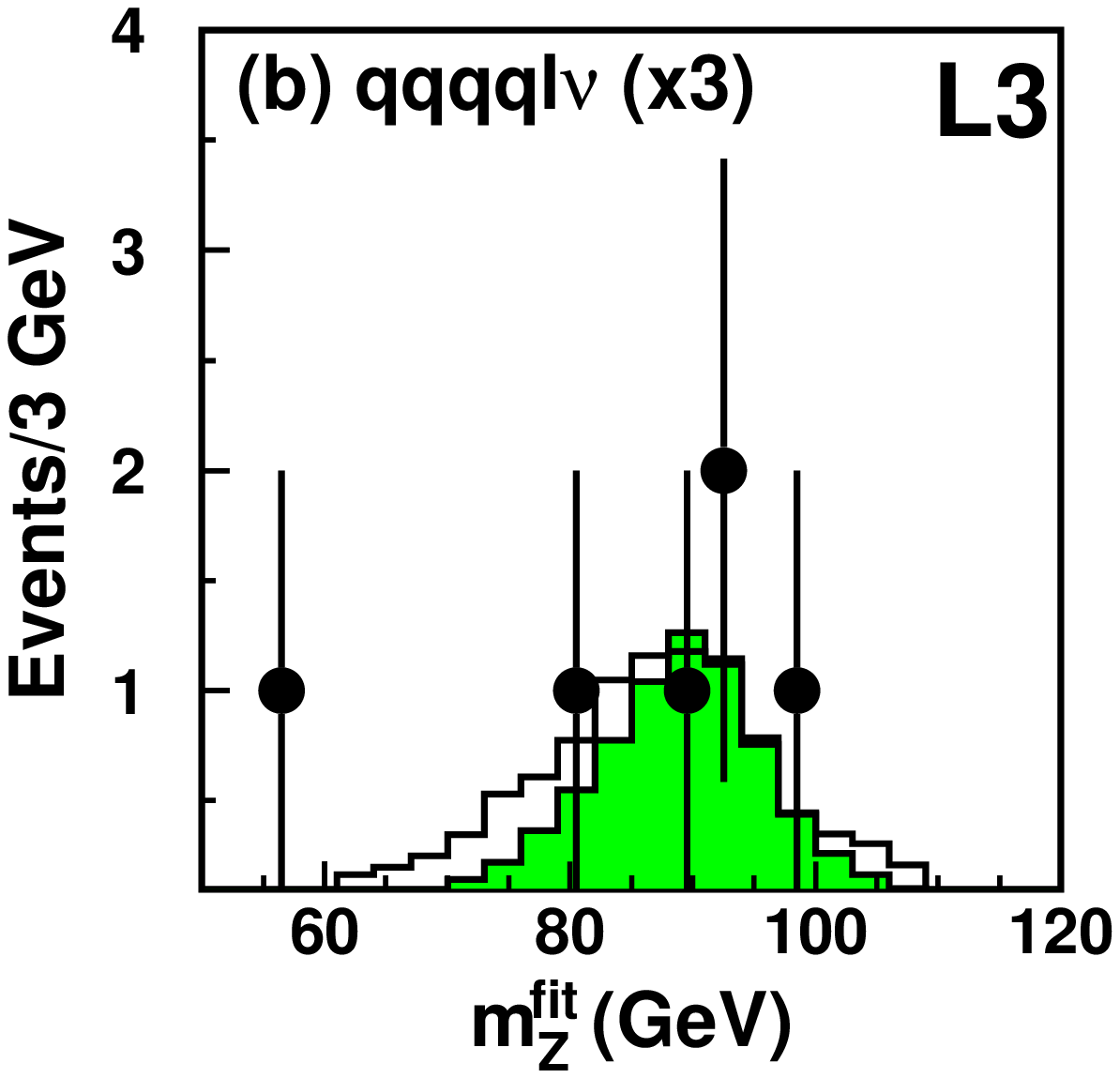} \\
\includegraphics[width=7.00cm]{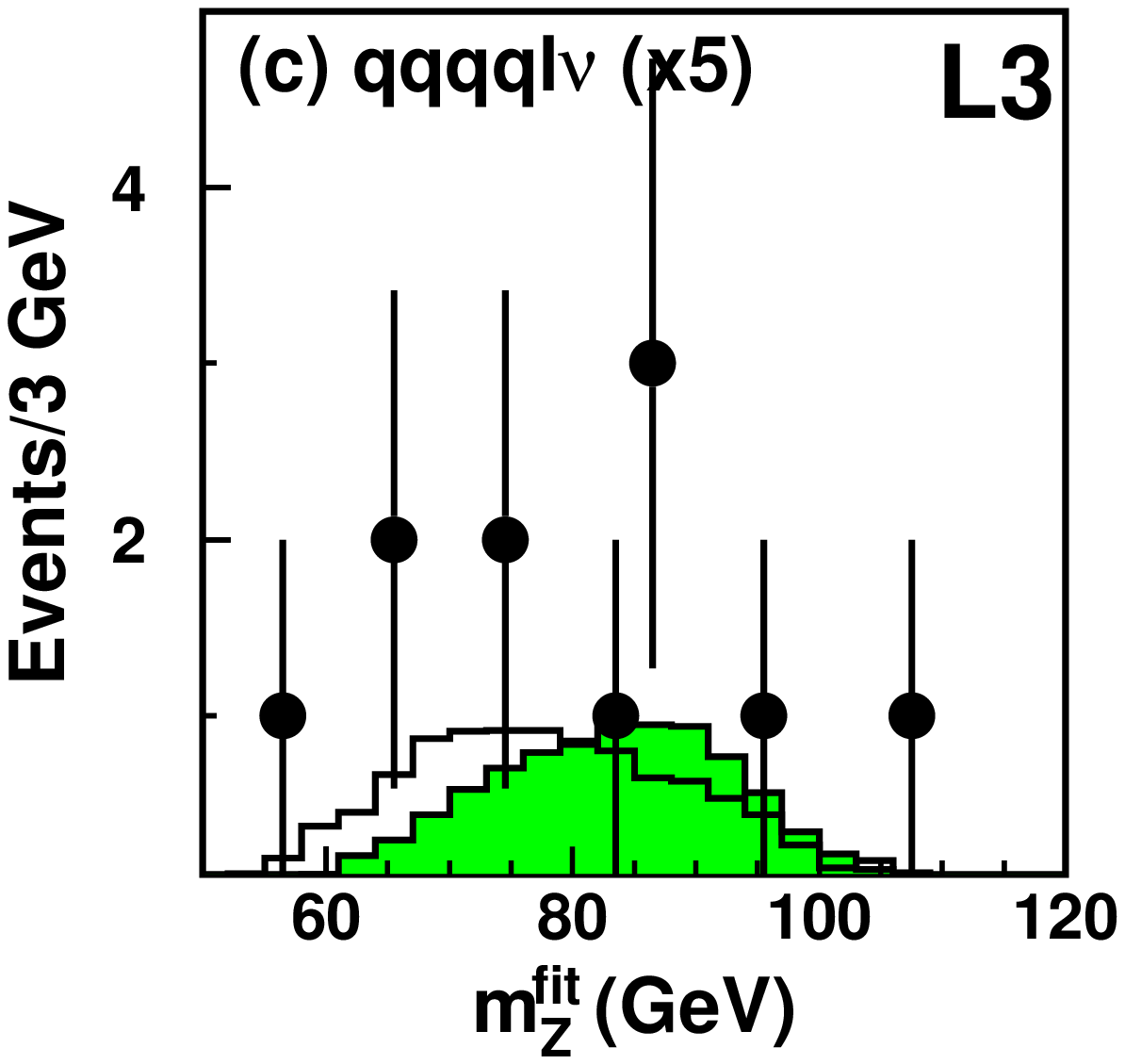} & \includegraphics[width=7.00cm]{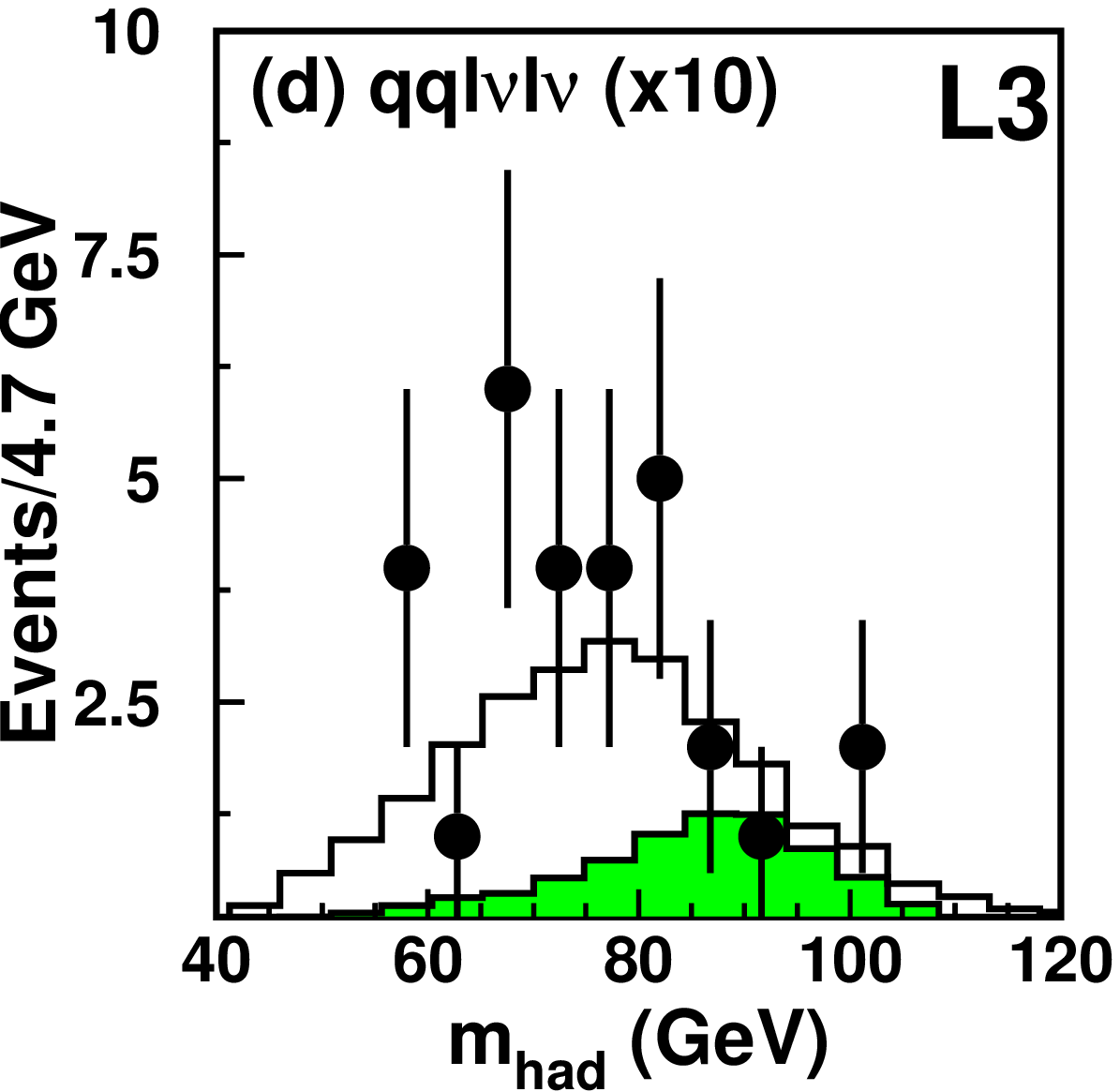} \\
\includegraphics[width=7.00cm]{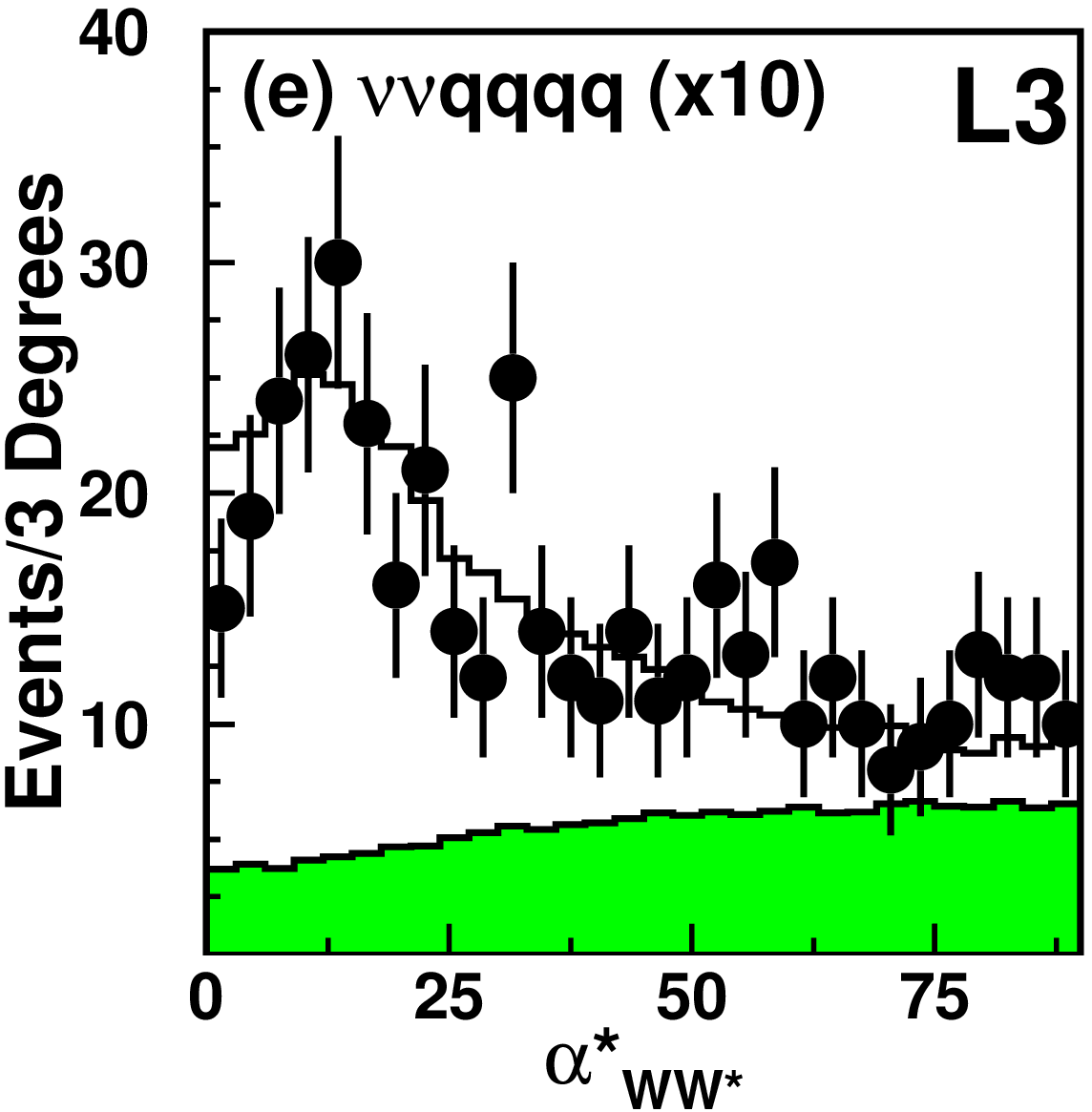} & \includegraphics[width=7.00cm]{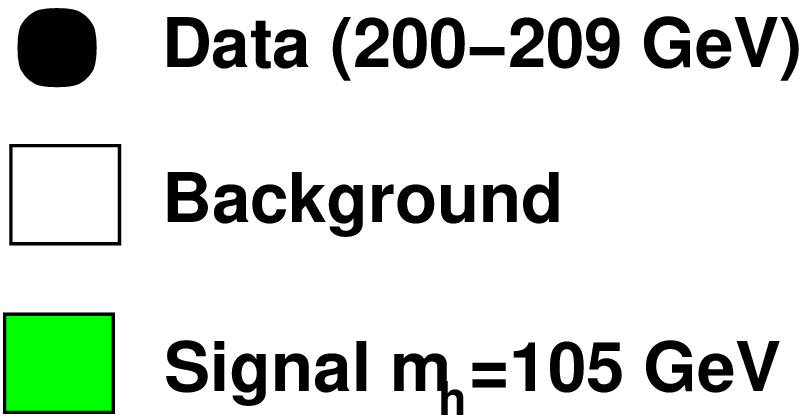}\\
\end{tabular}
\caption{
Distributions of variables used in input to the neural networks for data, background and 
signal Monte Carlo at $\mh=105\GeV$.  All centre-of-mass energies and 
subchannels are combined. b) refers to leptons originating from a W
boson while c) refers to leptons originating from a W$^*$ boson.
Signal Monte Carlo is magnified by the factor indicated in the figures.
}
\label{fig_input}
\end{center}
\end{figure}

  Two neural networks with ten input nodes are prepared for each lepton flavour to remove the
$\EE\to\WW$ and $\EE\to\mathrm{e}\nu\mathrm{qq}$ backgrounds. Besides
the common input variables $E^{max}_{jet}$, $E^{min}_{jet}$,
$\theta^{min}_{jj}$, $\chi^2_{\W\W}$, $m^{fit}_\Zo$ and $\log y_{34}$, the networks differ in the hypothesis used in associating the lepton-neutrino or quark-antiquark pairs to the W or the $\W^{(*)}$ bosons. For each network,
four additional variables, constructed after a kinematic fit to the
\qqqqlv\ topology, are used: the invariant masses of the
lepton-neutrino and quark-antiquark  candidate systems,
$m^{fit}_{\ell\nu}$ and $m^{fit}_{\rm qq}$, as well as their
respective momenta, $p^{fit}_{\ell\nu}$ and
$p^{fit}_{\rm qq}$. These momenta are expected to be small for the
signal hypothesis. Twenty hidden nodes and one output node are used
for each network.  The distributions in data, background and signal
Monte Carlo of the variable  $m^{fit}_\Zo$ used in the neural networks
are presented in Figures~\ref{fig_input}b and~\ref{fig_input}c for
the hypotheses of leptonic decays of the W and   $\W^{(*)}$ bosons, respectively.
The numbers of events expected and observed in
this channel after cuts on the neural network outputs are listed in
Table \ref{tab_4jlv}.

The mass of the Higgs boson is reconstructed from the recoil against
the two jets which form the Z candidate.  The final result of the
analysis is a discriminant variable combining the output of the neural
network and the reconstructed mass separately for each subchannel.
Figure~\ref{fig_fv}b shows the distributions of the final variable
with all six subchannels combined.

\subsection{\boldmath{$\mathrm{Zh}\to\mathrm{qq}\ell\nu\ell\nu$}}

In the $\mathrm{qq}\ell\nu\ell\nu$ channel, the Z decays into hadrons
while both W bosons decay into lepton-neutrino pairs.  The analysis
requires two identified leptons, one with more than 12$\GeV$ of energy
and the second with energy greater than 10$\GeV$.  The two neutrinos
in the signal signature imply a visible energy window between 50\% and
85\% of $\sqrt{s}$.  The analysis starts from events with hadronic
content and rejects the two-photon and $\EE\to\QQ(\gamma)$ backgrounds
as described above.

The dominant backgrounds after preselection are the $\EE\to\WW$ and
$\EE\to\mathrm{e}\nu \mathrm{qq}$ processes.  They are suppressed by
using a single neural network based on seven kinematic input
variables, with sixteen hidden nodes and one output node.  No
constrained fit is performed for the
$\mathrm{Zh}\to\mathrm{qq}\ell\nu\ell\nu$ hypothesis because the two
neutrinos render the technique ineffective, but some of the input
variables are similar to the ones employed in the previous sections:
$\log (y_{23}/y_{34})$, the hadronic invariant mass, $m_{had}$, which
should be consistent with $\MZ$, the minimum angles between each of
the leptons and the closest jet, $\theta^{min}_{lj}$, expected to be
small in the case of quark decays, the invariant masses of the
lepton-neutrino system and of the remaining of the event after a
kinematic fit to the $\WW\to\mathrm{qq}\ell\nu$ hypothesis,
$m^{fit}_{\ell\nu}$ and $m^{fit}_{rem}$, and the angle between
leptons, $\theta_{ll}$, which should be high in the background case.
Distributions of  $m_{had}$ in data, background and signal Monte
Carlo are presented in Figure~\ref{fig_input}d.
The numbers of events expected and observed in this channel after a
cut on the neural network output are listed in Table \ref{tab_qqlvlv}.

After selection, the events are divided into two groups: events where
neither lepton is identified as a tau and events where at least one of
the leptons is identified as a tau.  Most background events fall into
the second group. Final discriminant variables are constructed for
these two subchannels separately, using the reconstructed Higgs mass
and the neural network output. The Higgs mass is reconstructed by
scaling the jet masses and energies by a common factor until the dijet
mass is equal to \mZ, and then calculating the recoil mass against the
dijet.  Figure~\ref{fig_fv}c shows the distributions of the final
variable for data, background and signal, with the two groups of
events combined.

\subsection{\boldmath{$\mathrm{Zh}\to\nu\nu\mathrm{qqqq}$}}

In the case where the Z decays into two neutrinos and both W's decay
into hadrons, the signature is four jets plus missing energy. The
missing mass should be the Z mass and the visible mass the Higgs mass.
As in the $\mathrm{qqqqqq}$ channel, the $\mathrm{Zh}\rightarrow
\mathrm{ZZZ}^{*}\rightarrow \nu \nu \mathrm{qqqq}$ events are included
along with the \HWW\ signal.  The selection accepts events where
either the radiated Higgsstrahlung Z or the Z from the Higgs decays
into neutrinos, but not the $\mathrm{Z}^{*}\rightarrow \nu \nu$ case
because of insufficient missing energy. The accepted signatures
comprise 20$\%$ of the total $\mathrm{Zh}\rightarrow \mathrm{ZZZ}^*$
branching fraction, and their inclusion increases the expected signal
rate by 15\%. Most background comes from Z and W pair production
(especially $\EE\to\WW\rightarrow\mathrm{qq}\tau\nu$), and the
$\EE\to\QQ(\gamma)$ process.

The analysis starts from events with significant hadronic content as
well as substantial missing energy, suppressing $\EE\to\QQ(\gamma)$
and two-photon processes.  Events are constrained into a four-jet
topology using the Durham algorithm.  Events where any of the four
jets contains no charged track or has an energy less than 6 \GeV\ are
rejected.

The preselected events are subjected to a constrained kinematic fit
assuming a balanced event with four jets and an invisible Z.  The two
jets with the invariant mass closest to \mW\ are considered as the W
candidate dijet.  Three separate neural networks are created and used
to remove background processes.  These three networks, based on ten
input variables, twenty hidden nodes and one output node, are trained
against Z and W pair production and the $\EE\to\QQ(\gamma)$
process. 
The eight input variables used in the networks are: the recoil mass of the
event, $m_{rec}$, which should be consistent with $\MZ$ for signal events,
$E^{max}_{jet}$, $E^{min}_{jet}$, $\theta^{min}_{jj}$, $m^{fit}_{\W}$,
$\log y_{23}$, $\log y_{34}$ and $\alpha^*_{\W\W^*}$. The angle
$\alpha^*_{\W\W^*}$ corresponds to the angle $\alpha_{\W\W^*}$
calculated in the rest frame of the hadronic system.
Distributions in data, background and signal Monte Carlo of
$\alpha^*_{\W\W^*}$ are presented in Figure~\ref{fig_input}e.
Table \ref{tab_vv4j} gives the numbers of signal and
background events expected and data observed after cuts on the neural
network outputs.

The final variable is constructed by combining the outputs of the
three neural networks with the reconstructed Higgs mass in a single
discriminant.  Figure~\ref{fig_fv}d shows the distributions of this
final variable for data, background and signal.

\subsection{\boldmath{$\mathrm{Zh} \to \nu \nu \mathrm{qq}\ell\nu$}}

In the $\nu \nu \mathrm{qq}\ell\nu$ channel, the Z decays into
neutrinos and the $\mathrm{WW}^*$ pair decays into two quarks, a
lepton and a neutrino. The major backgrounds for this process depend
on lepton flavour and include W and Z pair production,
$\mathrm{e}\nu\mathrm{qq}$ and Zee final states, as well as hadronic
two-photon processes. As in the $\mathrm{qqqq}\ell\nu$ channel, the
signal divides into six subchannels as a function of the lepton
flavour and origin. The same variables and lepton identification
requirements are used to separate the subchannels.

The small visible energy in this channel makes the suppression of
two-photon processes particularly important. Stringent requirements
against this background are applied and events with moderate hadronic
content are selected.

After preselection, the most important background is
$\EE\to\WW\rightarrow\mathrm{qq}\ell\nu$ (particularly
$\mathrm{qq}\tau \nu$).  This background is rejected using several
variables including the average reconstructed W pair mass, $m_{\rm
W}^{ave}$. This variable is constructed by scaling the energies and
masses of the two jets by a common factor so that the sum of the jet
energy is $\sqrt{s}/2$, as would be the case in a
$\EE\to\WW\to\mathrm{qq}\ell\nu$ event.  Using these rescaled jets,
the W mass is calculated as the average of the dijet invariant mass
and the invariant mass of the lepton and the missing energy vector.

Each subchannel uses one neural network with eight input variables,
twenty hidden nodes and one output node in order to remove the \WW\
and $\mathrm{e}\nu \mathrm{qq}$ backgrounds. The neural network input
variables include $m_{\rm W}^{ave}$, the dijet invariant mass and the
missing invariant mass. In addition, most of the remaining kinematic
information in the event is used: the lepton energy, the hadronic
energy, the minimum angle between the lepton and jets, the angle
between the lepton and the dijet plane, and the angle between lepton
and missing energy vector.  The numbers of events predicted and
observed after cuts on the neural network outputs are listed in Table
\ref{tab_vvqqlv}.

There are insufficient constraints to fully reconstruct the Higgs
mass, so the visible mass is used as the final variable after a cut on
the neural network.  Figure~\ref{fig_fv}e shows the distributions of
the final variable with all six subchannels combined.

\subsection{\boldmath{$\mathrm{Zh}\to\ell\ell\mathrm{qqqq}$}}

In this channel, the Z decays into a pair of electrons or muons and
hadronic W decays are considered.  The event signature is two
energetic leptons with the invariant mass of the Z, two energetic jets
with an invariant mass near \mW\, and two lower energy jets. This
signature is also produced by 6.5\%\ of the \HZZ\ signal events.

The events are separated into two subchannels according to the lepton
flavour.  The analysis selects balanced energy events with significant
hadronic content.

The preselected events are fit to a topology with two leptons and four
jets and subjected to a kinematic fit requiring energy and momentum
conservation. The $\mathrm{eeqqqq}$ subchannel uses a neural network
which is trained against the $\EE\to\ZZ$ and $\EE\to\mathrm{Zee}$
backgrounds.  The $\mu \mu \mathrm{qqqq}$ subchannel uses two neural
networks, one trained to discriminate against the $\EE\to\ZZ$ process
and the other against $\EE\to\WW$.  The neural networks have six input
nodes, twenty five hidden nodes and one output variable.  The six input
variables are: $E^{max}_{jet}$, $E^{min}_{jet}$, $\theta^{min}_{jj}$,
$\theta^{min}_{lj}$, $\log Y_{34}$ and the invariant mass of the
dilepton system, which should be consistent with $\MZ$.  The number of
background and signal events expected and data events observed after
cuts on the neural network outputs are given in Table \ref{tab_ll4j}.

The final variable is a discriminant which combines the output of the
ZZ rejection neural network with the Higgs mass reconstructed from the
recoil against the lepton pair.  Figure~\ref{fig_fv}f shows the final
variable distributions with the two subchannels combined.

\section{Results}

\begin{figure}[H]
\begin{center}
\begin{tabular}{l}
\includegraphics[width=15cm]{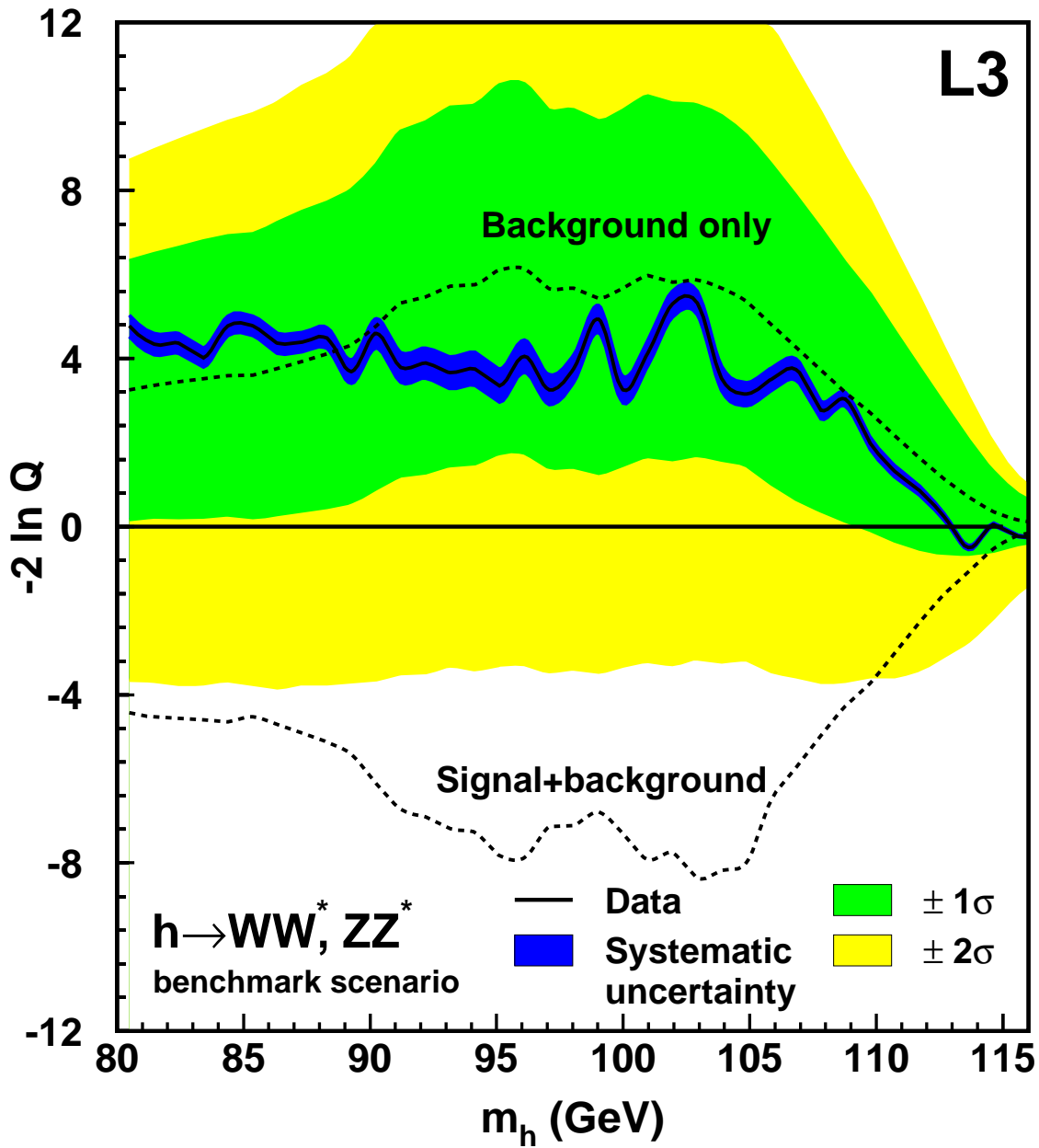} \\
\end{tabular}
\caption{
Log-likelihood ratio plot for the combined $\mathrm{h}\rightarrow \mathrm{WW}^*,\mathrm{ZZ}^*$ search. 
The dashed lines represent the value of
the expected background-only and signal+background distributions.  
The shaded regions around the background-only line indicate the $\pm
1\sigma$ and $\pm 2\sigma$ regions.  The solid line indicates the observed
values.  The bands around the observed sensitivity represent the effects of
systematic uncertainties.
}
\label{fig_llr}
\end{center}
\end{figure}

  The sensitivity to a Higgs boson decaying into weak boson pairs is
enhanced when all the previous analyses are combined. This combination
considers each final variable as an independent Poisson counting
experiment. A single statistical estimator from all channels is then
built. This statistical estimator is the log-likelihood
ratio~\cite{cl}:
\[ -2\ln{Q} = 2\sum_i s_i - 2\sum_i n_i \ln \left( 1+ \frac{s_i}{b_i} \right), \] 
where $n_i$ is the number of data events observed in the $i$-th bin of
the final variable, $s_i$ is the number of expected signal events and
$b_i$ is the number of expected background events.  The values of
$n_i$, $b_i$ and $s_i$ depend on the Higgs mass hypothesis.  The
$-2\ln{Q}$ values expected in the presence of background alone or both
signal and background are determined by replacing $n_i$ by $b_i$ or
$s_i+b_i$, respectively. The plot of the $-2\ln{Q}$ as a function of
the mass hypothesis is shown in Figure \ref{fig_llr}.  The dark band
surrounding the data gives the size of the systematic uncertainties.
The systematic dependence of the $-2\ln{Q}$ distribution on detector
quantities such as calorimeter energy scale and tracking efficiency is
determined by shifting correlated variables in the analysis and
propagating the effects to the signal and background hypothesis
distributions.  Other factors, such as signal and background Monte
Carlo statistics and cross section uncertainties, are also included.
Table~\ref{tab_syst} details the different sources of systematic and
their contribution to the total systematic uncertainty on $-2\ln{Q}$. 

\begin{figure}[H]
\begin{center}
\begin{tabular}{l}
\includegraphics[width=15cm]{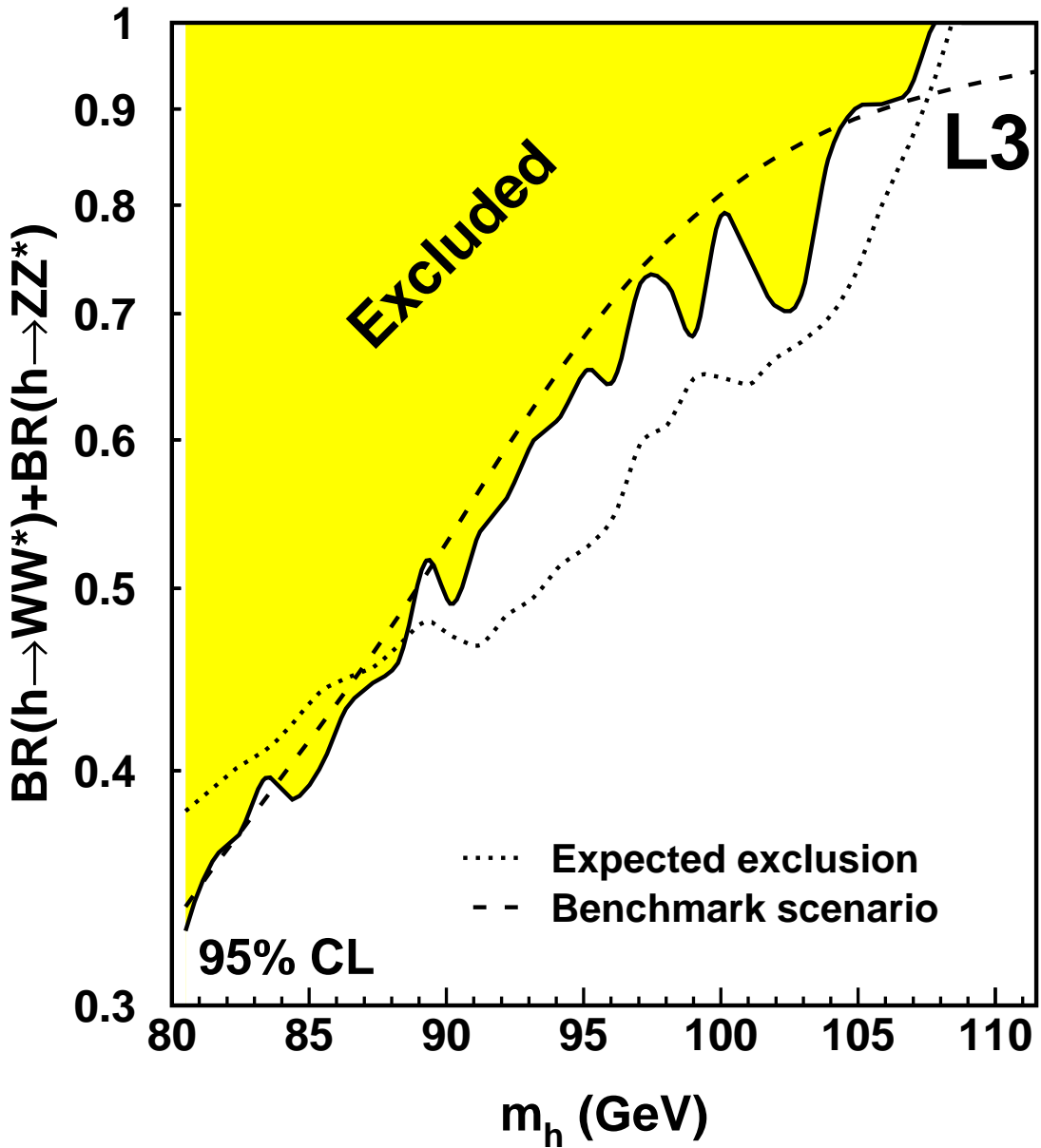} \\
\end{tabular}
\caption{Excluded values at 95$\%$ CL of $\mathrm{BR}(\HWW)+\mathrm{BR}(\HZZ)$ as 
a function of the Higgs mass (solid line), in the assumption of the Standard 
Model production cross section. The expected 95$\%$ CL limit 
(dashed line) and the fermiophobic benchmark prediction (dotted line) are also presented.
The Standard Model prediction for $\mathrm{BR}(\HWW)$ is 8\% at $\mh=115\GeV$,
falling below 1\% for $\mh<100\GeV$.
}
\label{fig_br}
\end{center}
\end{figure}

No significant excess indicating the presence of a Higgs boson
decaying into WW$^*$ or ZZ$^*$ is observed in the data, which would
manifest as a significant dip in the $-2\ln{Q}$ distribution.
Confidence levels for the absence of a signal are hence derived
\cite{cl}.  Figure \ref{fig_br} shows the observed and expected 95\%
confidence level (CL) limits as a function of the Higgs mass, assuming
the Standard Model production cross section.  In the assumption of
BR(\htoWW)+BR(\htoZZ)=100\%, a Higgs boson with mass less than $108.1
\GeV$ is excluded at 95\% CL.  Assuming the value of
BR(\htoWW)+BR(\htoZZ) of the benchmark fermiophobic scenario,
calculated with the HDECAY program \cite{hdecay}, the observed
exclusion region is
$\mathrm{83.7}\GeV\mathrm{<}\mh\mathrm{<104.6}\GeV$ with a region
between $\mathrm{88.9}\GeV\mathrm{<}\mh\mathrm{<89.4}\GeV$ which can
be excluded only at 93\% CL.

\begin{figure}[H]
\begin{center}
\begin{tabular}{l}
\includegraphics[width=15cm]{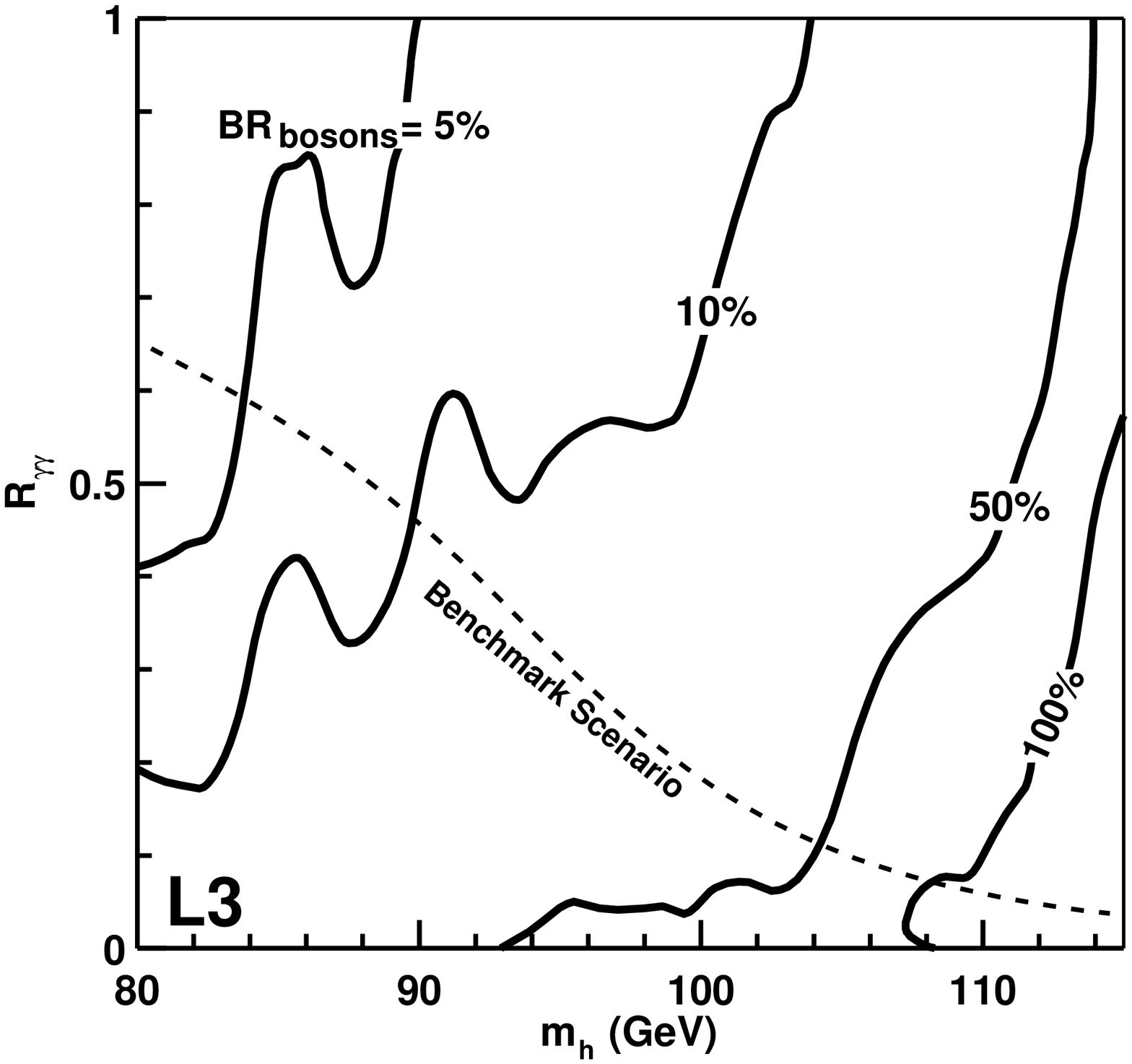} \\
\end{tabular}
\caption{
  The 95\% CL limit for \Brphobic\ as a function of \mh\ and \Brgg.
The solid lines indicate the borders between regions of exclusion.
The crossing point 
between the $\Brphobic=100\%$ line and the dashed line provides the lower limit 
on the Higgs mass in the benchmark scenario: $\mh > 108.3 \GeV$. The minimum 
value of $\mh$ on the $\Brphobic=100\%$ contour gives a lower mass limit 
for any model where the Higgs decays exclusively into electroweak boson pairs:
$\mh > 107 \GeV$. 
}
\label{fig_scan}
\end{center}
\end{figure}

Model independent fermiophobic results can be derived by scanning the
relative branching fractions of \htogg\ and \htoWW.  The branching
fractions into boson pairs can be conveniently parameterised in the
form:
\begin{eqnarray*}
\Brphobic & = &
\mathrm{BR}(\htogg)~+~\mathrm{BR}(\htoWW)~+~\mathrm{BR}(\htoZZ), \\
\Brgg &= &\mathrm{BR}(\htogg)~/~\Brphobic.
\end{eqnarray*} 
 \Brgg\ represents the fraction of fermiophobic decays into photon
pairs, and ranges from zero to one, while \Brphobic\ represents the
total Higgs branching fraction to pairs of gauge bosons.  The scan
combines these \htoWW\ results with the previously published L3
\htogg\ results \cite{l3ggpaper}, determining the 95\% CL exclusion
for \Brphobic\ at each point in the \mh\ versus \Brgg\ plane.  The
full scan results are presented in Figure~\ref{fig_scan}.  In the
benchmark scenario, the fermiophobic Higgs is limited at 95\% CL to
have $\mh > 108.3 \GeV$, compared to an expected limit of $\mh > 110.7
\GeV$. These results represent a significant extension of the $\mh >
105.4 \GeV$ limit obtained in the photonic channel~\cite{l3ggpaper}.
These results also exclude at 95\% CL any model with $\mh < 107 \GeV$
for $\Brphobic=100\%$ and any value of \Brgg, assuming the Standard
Model production cross section.

%%%%%%%%%%%%%%%%%%%%%%%%%%%%%%%%%%%%%%%%%%%%%%%%%%%%%%%%%%%
% References
\newpage
\bibliographystyle{l3style}

%%%%%%%%%%%%%%%%%%%%%%%%%%%%%%%%%%%%%%%%%%%%%%%%%%%%%%%%%%%
% Author list
\newpage
\input namelist261.tex

%%%%%%%%%%%%%%%%%%%%%%%%%%%%%%%%%%%%%%%%%%%%%%%%%%%%%%%%%%%
% Tables
\newpage

%*******************************************************
%***       table
%*******************************************************
\newpage
%*******************************************************
%***       lumi table
%*******************************************************
\begin{table} [ht]
\begin{center}
\hspace*{-1.cm}
\begin{tabular}{|c|c c c c c c|}
\hline
$\sqrt s$ (GeV) &199.5 & 201.8 & 203.1& 205.0& 206.5& 208.0 \\
\hline
Luminosity (pb$^{-1}$) &\pho82.8&\pho37.0&\pho\pho8.8&\pho68.9&130.4&\pho\pho8.5 \\
\hline
\end{tabular}
\caption{
Average centre-of-mass energies and corresponding integrated luminosities.
}
\label{tab_lumi}
\end{center}
\end{table}
%*******************************************************
%***       6j table
%*******************************************************
\begin{table} [ht]
\begin{center}
\hspace*{-1.cm}
\begin{tabular}{|c|c|c||c|c|c||c|}
\hline
$\mathrm{Zh}\rightarrow\mathrm{qqqqqq}$ 
             & $\mathrm{Data}$ & Background & $\WW$ & $\mathrm{ZZ}$ & $\QQ(\gamma)$ &  Signal \\ \hline
Preselection & 1886 & 1870.1 & 1274.4 & 104.6 & 488.9 & 16.6 \\ \hline
Selection    &\pho443 &\pho446.0 &\pho347.7 &\pho  44.1 &\pho54.0 & 14.4 \\ 
\hline
\end{tabular}
\caption{
Number of events observed in data by the 
$\mathrm{Zh}\rightarrow\mathrm{qqqqqq}$ analysis, compared with the Standard 
Model expectations.  The Monte Carlo breakdown in different processes is 
given. Signal expectations are given for $\mh = 105 \GeV$
in the fermiophobic benchmark scenario.
}
\label{tab_6j}
\end{center}
\end{table}%%
%*******************************************************
%***       4jlv table
%*******************************************************
\begin{table} [ht]
\begin{center}
\hspace*{-1.cm}
\begin{tabular}{|c|c|c||c|c|c||c|}
\hline
$\mathrm{Zh}\rightarrow\mathrm{qqqqe}\nu$
                  & Data & Background & \WW\   & ZZ  & $\QQ(\gamma)$ & Signal \\ \hline
Preselection      &   13 &       12.5 &    4.1 & 2.3 &           2.8 & 2.3 \\ \hline
Selection         &\pho6 &    \pho5.3 &    1.5 & 1.4 &           1.1 & 2.0 \\ \hline
\hline

$\mathrm{Zh}\rightarrow\mathrm{qqqq}\mu\nu$
                  & Data & Background & \WW\ & ZZ  & $\QQ(\gamma)$ & Signal \\ \hline
Preselection      &    3 &        7.3 &  5.0 & 1.4 &           0.9 & 1.7 \\ \hline
Selection         &    1 &        3.0 &  1.7 & 0.8 &           0.5 & 1.5 \\ \hline
\hline
$\mathrm{Zh}\rightarrow\mathrm{qqqq}\tau\nu$
                  &  Data & Background &    \WW\ &   ZZ   & $\QQ(\gamma)$ & Signal \\ \hline
Preselection      &   251 &      202.4 &   137.5 &   13.9 &          43.4 & 1.6 \\ \hline
Selection         &\pho41 &   \pho41.2 &\pho24.4 &\pho2.8 &       \pho8.6 &1.3 \\ 
\hline
\end{tabular}
\caption{
Number of events observed in data by the $\mathrm{Zh}\rightarrow\mathrm{qqqq}\ell\nu$ analysis, compared with the Standard Model expectations.  The Monte Carlo breakdown in different processes is given. Signal
expectations are given for $\mh = 105 \GeV$
in the fermiophobic benchmark scenario.
}
\label{tab_4jlv}
\end{center}
\end{table}%%
%*******************************************************
%***       qqlvlv table
%*******************************************************
\begin{table} [ht]
\begin{center}
\hspace*{-1.cm}
\begin{tabular}{|c|c|c||c|c|c||c|}
\hline
$\mathrm{Zh}\rightarrow\mathrm{qq}\ell\nu\ell\nu$
                  &   Data & Background &   \WW &  ZZ & $\mathrm{e}\nu\mathrm{qq}$ & Signal \\ \hline
Preselection      &     23 &       21.4 &   10.7 & 1.6 &                7.6 & 0.7 \\\hline
Selection         &  \pho2 &    \pho1.4 &\pho0.5 & 0.5 &                0.2 & 0.5 \\ \hline
\hline
$\mathrm{Zh}\rightarrow\mathrm{qq}\tau\nu\ell\nu$
                  &   Data & Background &   \WW &  ZZ & $\mathrm{e}\nu\mathrm{qq}$ & Signal \\ \hline
Preselection      &     96 &       91.7 &  57.3 & 6.4 &               20.8 & 0.6 \\\hline
Selection         &     28 &       23.1 &  13.1 & 3.5 &            \pho4.6 & 0.4 \\\hline
\end{tabular}
\caption{
Number of events observed in data by the $\mathrm{Zh}\rightarrow\mathrm{qq}\ell\nu\ell\nu$ analysis, compared with the Standard Model expectations.  The Monte Carlo breakdown in different processes is given. 
Signal expectations are given for $\mh = 105 \GeV$
in the fermiophobic benchmark scenario.
}
\label{tab_qqlvlv}\end{center}\end{table}
%*******************************************************
%***       vv4j table
%*******************************************************
\begin{table} [ht]
\begin{center}
\hspace*{-1.cm}
\begin{tabular}{|c|c|c||c|c|c|c||c|}\hline
$\mathrm{Zh}\rightarrow\nu\nu\mathrm{qqqq}$
                  &   Data & Background &     \WW &    ZZ  & $\mathrm{e}\nu\mathrm{qq}$ & $\QQ(\gamma)$ & Signal \\ \hline
Preselection      &    451 &      439.7 &   263.2 &   25.3 &               73.6 &          75.9 & 4.0 \\\hline
Selection         & \pho41 &   \pho47.9 &\pho41.5 &\pho4.9 &            \pho4.4 &       \pho6.9 & 3.3 \\ \hline
\end{tabular}
\caption{
Number of events observed in data by the 
$\mathrm{Zh}\rightarrow\nu\nu\mathrm{qqqq}$ analysis, compared with the 
Standard Model expectations.  The Monte Carlo breakdown in different 
processes is given.  Signal expectations are given for $\mh = 105 \GeV$
in the fermiophobic benchmark scenario.
}
\label{tab_vv4j}
\end{center}
\end{table}%%
%*******************************************************
%***       vvqqlv table
%*******************************************************
\begin{table} [ht]
\begin{center}
\hspace*{-1.cm}
\begin{tabular}{|c|c|c||c|c|c||c|}
\hline
$\mathrm{Zh}\rightarrow\nu\nu\mathrm{qq}\ell\nu$
                  &   Data & Background &   \WW &  ZZ & $\mathrm{e}\nu\mathrm{qq}$ & Signal \\ \hline
Preselection      &     21 &       11.3 &   5.7 & 1.2 &                3.9 & 1.0 \\ \hline
Selection         &  \pho2 &    \pho1.7 &   1.0 &  -  &                0.7 & 0.8 \\ \hline
\hline
$\mathrm{Zh}\rightarrow\nu\nu\mathrm{qq}(\ell\nu)^*$
                  &   Data & Background &   \WW  &  ZZ & $\mathrm{e}\nu\mathrm{qq}$ & Signal \\ \hline
Preselection      &     39 &       44.4 &   34.4 & 3.6 &                5.4 & 0.6\\ \hline
Selection         &  \pho7 &    \pho5.3 &\pho3.9 & 0.5 &                0.9 & 0.6\\ \hline
\end{tabular}
\caption{
Number of events observed in data by the $\mathrm{Zh}\rightarrow\nu\nu\mathrm{qq}\ell\nu$ analysis, compared with the Standard Model expectations.  The Monte Carlo breakdown in different processes is given. 
Signal expectations are given for $\mh = 105 \GeV$
in the fermiophobic benchmark scenario.
}
\label{tab_vvqqlv}
\end{center}
\end{table}%%
%*******************************************************
%***       ll4j table
%*******************************************************
\begin{table} [ht]
\begin{center}
\hspace*{-1.cm}
\begin{tabular}{|c|c|c||c|c||c|}
\hline
$\mathrm{Zh}\rightarrow\mathrm{eeqqqq}$
                  & Data & Background & Zee & ZZ  & Signal \\ \hline
Preselection      &    4 &        6.8 & 1.3 & 5.3 & 0.4 \\ \hline
Selection         &    2 &        3.0 & 0.7 & 2.3 & 0.4 \\ \hline
\hline
$\mathrm{Zh}\rightarrow\mu\mu\mathrm{qqqq}$
                  & Data & Background & \WW & ZZ  & Signal \\ \hline
\hline
Preselection      &   13 &       13.7 & 7.2 & 5.5 & 0.4 \\ \hline
Selection         &\pho3 &    \pho4.2 & 1.4 & 2.7 & 0.4 \\ \hline
\end{tabular}
\caption{
Number of events observed in data by the $\mathrm{Zh}\rightarrow\ell\ell\mathrm{qqqq}$ analysis, compared with the Standard Model expectations.  The Monte Carlo breakdown in different processes is given. 
Signal expectations are given for $\mh = 105 \GeV$
in the fermiophobic benchmark scenario.
}
\label{tab_ll4j}
\end{center}
\end{table}

%*******************************************************
%***       syst table
%*******************************************************
\begin{table} [ht]
\begin{center}
\hspace*{-1.cm}
\begin{tabular}{|l|c|c|c|c|c|c|}
\hline
                         & \multicolumn{6}{|l|}{Analysis $\mathrm{Zh\rightarrow}$} \\
\hline
Source of Systematics    & $\mathrm{qqqqqq}$ & $\mathrm{qqqq\ell\nu}$ 
                         & $\mathrm{qq\ell\nu\ell\nu}$ & $\mathrm{\nu\nu qqqq}$ 
                         & $\mathrm{\nu\nu qq\ell\nu}$ & $\mathrm{\ell\ell qqqq}$ \\
\hline
Tracking effciency                  & 2.1\% & 1.7\% & 0.4\% & 0.2\% & 1.8\% & 2.1\% \\
Energy scale                        & 1.3\% & 2.9\% & 2.0\% & 6.0\% & 5.3\% & 2.2\% \\
Lepton identification               & --    & 3.8\% & 2.8\% & --    & 4.4\% & 1.8\% \\
Event shape                         & 1.2\% &  --   & --    & 2.6\% & --    & --    \\
Signal Monte Carlo statistics       & 2.0\% & 3.0\% & 4.0\% & 2.0\% & 2.0\% & 2.0\% \\
Background Monte Carlo statistics   & 0.1\% & 1.2\% & 0.1\% & 0.1\% & 1.1\% & 0.1\% \\
Background cross sections           & 0.4\% & 1.2\% & 0.8\% & 0.7\% & 1.1\% & 0.4\% \\
\hline
Total                               & 3.4\% & 6.1\% & 5.3\% & 6.9\% & 7.6\% & 4.1\% \\
\hline
\end{tabular}
\caption{Systematic uncertainties on the  $-2\ln{Q}$ distributions for the different channels.}
\label{tab_syst}
\end{center}
\end{table}

%*******************************************************
%***       figures 
%*******************************************************
%
%**********************************************************************

\newpage

\newpage

\newpage

\newpage

\newpage

\newpage

%  
%%%%%%%%%%%%%%%%%%%%%%%%%%%%%%%%%%%%%%%%%%%%%%%%%%%%%%%%%%%%%%%%%%%%%%%%%
\vfill\newpage
\end{document}

%% file: namelist261.tex
\typeout{   }     
\typeout{Using author list for paper 261 -  }
\typeout{$Modified: Jul 15 2001 by smele $}
\typeout{!!!!  This should only be used with document option a4p!!!!}
\typeout{   }
%
%
%
%  L A T E X  version!!
%
%
% Make sure that the Lep package has been used!
%\input{Lep.sty}%
%
%\ifx\LepCalled\undefined%
%\typeout{     }%
%\typeout{!!!!!!!!!!!!!!!!!!!!!!!!!!!!!!!!!!!!!!!!!!!!!!!!!!!!!!!!!!!}%
%\typeout{Yikes.  You haven't used the Lep package!}%
%\typeout{Please put \protect\usepackage\protect{Lep\protect} in your preamble,
%         followed by}%
%\typeout{\protect\Lep\protect{1\protect} or \protect\Lep\protect{2\protect}}%
%\typeout{     }%
%\typeout{For now you will get a Lep phase 2 authorlist (may not be right!).}%
%\typeout{!!!!!!!!!!!!!!!!!!!!!!!!!!!!!!!!!!!!!!!!!!!!!!!!!!!!!!!!!!!}%
%\typeout{     }%
%\Lep{2}\fi%

\newcount\tutecount  \tutecount=0
\def\tutenum#1{\global\advance\tutecount by 1 \xdef#1{\the\tutecount}}
\def\tute#1{$^{#1}$}
\tutenum\aachen            % 1
\tutenum\nikhef            % 2
\tutenum\mich              % 3
\tutenum\lapp              % 4
\tutenum\basel             % 5
\tutenum\lsu               % 6
\tutenum\beijing           % 7
\tutenum\bologna           % 8
\tutenum\tata              % 9 
\tutenum\ne                % 10
\tutenum\bucharest         % 11
\tutenum\budapest          % 12
\tutenum\mit               % 13
\tutenum\panjab            % 14 
\tutenum\debrecen          % 15
\tutenum\dublin            % 16
\tutenum\florence          % 17
\tutenum\cern              % 18
\tutenum\wl                % 19
\tutenum\geneva            % 20
\tutenum\hefei             % 21
\tutenum\lausanne          % 22
\tutenum\lyon              % 23
\tutenum\madrid            % 24
\tutenum\florida           % 25
\tutenum\milan             % 26
\tutenum\moscow            % 27
\tutenum\naples            % 29
\tutenum\cyprus            % 30
\tutenum\nymegen           % 31
\tutenum\caltech           % 32
\tutenum\perugia           % 33
\tutenum\peters            % 34
\tutenum\cmu               % 35
\tutenum\potenza           % 36
\tutenum\prince            % 37
\tutenum\riverside         % 38
\tutenum\rome              % 39
\tutenum\salerno           % 40
\tutenum\ucsd              % 41
\tutenum\sofia             % 42
\tutenum\korea             % 43
\tutenum\purdue            % 44
\tutenum\psinst            % 45
\tutenum\zeuthen           % 46
\tutenum\eth               % 47
\tutenum\hamburg           % 48
\tutenum\taiwan            % 49
\tutenum\tsinghua          % 50

{
\parskip=0pt
\noindent
{\bf The L3 Collaboration:}
\ifx\selectfont\undefined%  old style font selection
 \baselineskip=10.8pt
 \baselineskip\baselinestretch\baselineskip
 \normalbaselineskip\baselineskip
 \ixpt
\else%                      new style font selection
 \fontsize{9}{10.8pt}\selectfont
\fi
\medskip
\tolerance=10000
\hbadness=5000
\raggedright
\hsize=162truemm\hoffset=0mm
\def\r{\rlap,}
\noindent

P.Achard\r\tute\geneva\ 
O.Adriani\r\tute{\florence}\ 
M.Aguilar-Benitez\r\tute\madrid\ 
J.Alcaraz\r\tute{\madrid,\cern}\ 
G.Alemanni\r\tute\lausanne\
J.Allaby\r\tute\cern\
A.Aloisio\r\tute\naples\ 
M.G.Alviggi\r\tute\naples\
H.Anderhub\r\tute\eth\ 
V.P.Andreev\r\tute{\lsu,\peters}\
% ...
\newpage
%\rule{\textwidth}{0.4pt}
\begin{list}{A}{\itemsep=0pt plus 0pt minus 0pt\parsep=0pt plus 0pt minus 0pt
                \topsep=0pt plus 0pt minus 0pt}
\item[\aachen]
 III. Physikalisches Institut, RWTH, D-52056 Aachen, Germany$^{\S}$
\item[\nikhef] National Institute for High Energy Physics, NIKHEF, 
     and University of Amsterdam, NL-1009 DB Amsterdam, The Netherlands
\item[\mich] University of Michigan, Ann Arbor, MI 48109, USA
\item[\lapp] Laboratoire d'Annecy-le-Vieux de Physique des Particules, 
     LAPP,IN2P3-CNRS, BP 110, F-74941 Annecy-le-Vieux CEDEX, France
\item[\basel] Institute of Physics, University of Basel, CH-4056 Basel,
     Switzerland
\item[\lsu] Louisiana State University, Baton Rouge, LA 70803, USA
\item[\beijing] Institute of High Energy Physics, IHEP, 
  100039 Beijing, China$^{\triangle}$ 
\item[\bologna] University of Bologna and INFN-Sezione di Bologna, 
     I-40126 Bologna, Italy
\item[\tata] Tata Institute of Fundamental Research, Mumbai (Bombay) 400 005, India
\item[\ne] Northeastern University, Boston, MA 02115, USA
\item[\bucharest] Institute of Atomic Physics and University of Bucharest,
     R-76900 Bucharest, Romania
\item[\budapest] Central Research Institute for Physics of the 
     Hungarian Academy of Sciences, H-1525 Budapest 114, Hungary$^{\ddag}$
\item[\mit] Massachusetts Institute of Technology, Cambridge, MA 02139, USA
\item[\panjab] Panjab University, Chandigarh 160 014, India.
\item[\debrecen] KLTE-ATOMKI, H-4010 Debrecen, Hungary$^\P$
\item[\dublin] Department of Experimental Physics,
  University College Dublin, Belfield, Dublin 4, Ireland
\item[\florence] INFN Sezione di Firenze and University of Florence, 
     I-50125 Florence, Italy
\item[\cern] European Laboratory for Particle Physics, CERN, 
     CH-1211 Geneva 23, Switzerland
\item[\wl] World Laboratory, FBLJA  Project, CH-1211 Geneva 23, Switzerland
\item[\geneva] University of Geneva, CH-1211 Geneva 4, Switzerland
\item[\hefei] Chinese University of Science and Technology, USTC,
      Hefei, Anhui 230 029, China$^{\triangle}$
\item[\lausanne] University of Lausanne, CH-1015 Lausanne, Switzerland
\item[\lyon] Institut de Physique Nucl\'eaire de Lyon, 
     IN2P3-CNRS,Universit\'e Claude Bernard, 
     F-69622 Villeurbanne, France
\item[\madrid] Centro de Investigaciones Energ{\'e}ticas, 
     Medioambientales y Tecnol\'ogicas, CIEMAT, E-28040 Madrid,
     Spain${\flat}$ 
\item[\florida] Florida Institute of Technology, Melbourne, FL 32901, USA
\item[\milan] INFN-Sezione di Milano, I-20133 Milan, Italy
\item[\moscow] Institute of Theoretical and Experimental Physics, ITEP, 
     Moscow, Russia
\item[\naples] INFN-Sezione di Napoli and University of Naples, 
     I-80125 Naples, Italy
\item[\cyprus] Department of Physics, University of Cyprus,
     Nicosia, Cyprus
\item[\nymegen] University of Nijmegen and NIKHEF, 
     NL-6525 ED Nijmegen, The Netherlands
\item[\caltech] California Institute of Technology, Pasadena, CA 91125, USA
\item[\perugia] INFN-Sezione di Perugia and Universit\`a Degli 
     Studi di Perugia, I-06100 Perugia, Italy   
\item[\peters] Nuclear Physics Institute, St. Petersburg, Russia
\item[\cmu] Carnegie Mellon University, Pittsburgh, PA 15213, USA
\item[\potenza] INFN-Sezione di Napoli and University of Potenza, 
     I-85100 Potenza, Italy
\item[\prince] Princeton University, Princeton, NJ 08544, USA
\item[\riverside] University of Californa, Riverside, CA 92521, USA
\item[\rome] INFN-Sezione di Roma and University of Rome, ``La Sapienza",
     I-00185 Rome, Italy
\item[\salerno] University and INFN, Salerno, I-84100 Salerno, Italy
\item[\ucsd] University of California, San Diego, CA 92093, USA
\item[\sofia] Bulgarian Academy of Sciences, Central Lab.~of 
     Mechatronics and Instrumentation, BU-1113 Sofia, Bulgaria
\item[\korea]  The Center for High Energy Physics, 
     Kyungpook National University, 702-701 Taegu, Republic of Korea
\item[\purdue] Purdue University, West Lafayette, IN 47907, USA
\item[\psinst] Paul Scherrer Institut, PSI, CH-5232 Villigen, Switzerland
\item[\zeuthen] DESY, D-15738 Zeuthen, Germany
\item[\eth] Eidgen\"ossische Technische Hochschule, ETH Z\"urich,
     CH-8093 Z\"urich, Switzerland
\item[\hamburg] University of Hamburg, D-22761 Hamburg, Germany
\item[\taiwan] National Central University, Chung-Li, Taiwan, China
\item[\tsinghua] Department of Physics, National Tsing Hua University,
      Taiwan, China
\item[\S]  Supported by the German Bundesministerium 
        f\"ur Bildung, Wissenschaft, Forschung und Technologie
\item[\ddag] Supported by the Hungarian OTKA fund under contract
numbers T019181, F023259 and T037350.
\item[\P] Also supported by the Hungarian OTKA fund under contract
  number T026178.
\item[$\flat$] Supported also by the Comisi\'on Interministerial de Ciencia y 
        Tecnolog{\'\i}a.
\item[$\sharp$] Also supported by CONICET and Universidad Nacional de La Plata,
        CC 67, 1900 La Plata, Argentina.
\item[$\triangle$] Supported by the National Natural Science
  Foundation of China.
\end{list}
}
\vfill

%%% Local Variables: 
%%% mode: latex
%%% TeX-master: t
%%% End: